\documentclass[10pt,preprint,a4paper]{aastex}
\usepackage{amsmath}                
\usepackage{amsfonts}               
\usepackage{amssymb}                
\usepackage{epsfig}                 

\newcommand{\kms}{{~\rm km\; s^{-1}}}

\newcommand{\cm}{{~\rm cm}}
\newcommand{\km}{{~\rm km}}
\newcommand{\s}{{~\rm s}}

\newcommand{\K}{{~\rm K}}
\newcommand{\erg}{{~\rm erg}}
\newcommand{\yr}{{~\rm yr}}

\newcommand{\Gyr}{{~\rm Gyr}}
\newcommand{\pc}{{~\rm pc}}
\newcommand{\kpc}{{~\rm kpc}}

\newcommand{\AU}{{~\rm AU}}

\def \astrobj#1{#1}

\begin{document}

\title{THE JET FEEDBACK MECHANISM (JFM) IN STARS, GALAXIES AND CLUSTERS}

\author{Noam Soker\altaffilmark{1}}

\altaffiltext{1}{Department of Physics, Technion -- Israel Institute of Technology, Haifa
32000, Israel; soker@physics.technion.ac.il}

\begin{abstract}
I review the influence jets and the bubbles they inflate might have on their ambient gas as they operate through a negative jet feedback mechanism (JFM). I discuss astrophysical systems where jets are observed to influence the ambient gas, in many cases by inflating large, hot, and low-density bubbles, and systems where the operation of the JFM is still a theoretical suggestion. The first group includes cooling flows in galaxies and clusters of galaxies, star-forming galaxies, young stellar objects, and bipolar planetary nebulae. The second group includes core collapse supernovae, the common envelope evolution, the grazing envelope evolution, and intermediate luminosity optical transients. The suggestion that the JFM operates in these four types of systems is based on the assumption that jets are much more common than what is inferred from objects where they are directly observed.
Common to all eight types of systems reviewed here is the presence of a compact object inside an extended ambient gas. The ambient gas serves as a potential reservoir of mass to be accreted on to the compact object. If the compact object launches jets as it accretes mass, the jets might reduce the accretion rate as they deposit energy to the ambient gas, or even remove the entire ambient gas, hence closing a negative feedback cycle.
\end{abstract}


\section{INTRODUCTION}
\label{sec:intro}

Large varieties of astrophysical objects are composed of a compact object that resides inside a more massive and much more extended ambient gas. This review deals with the manner by which the compact object influences the evolution of the ambient gas by launching jets, and the jets' operation is regulated by a negative feedback mechanism. This is termed the jet-feedback mechanism (JFM).
This review does not deal with jet-launching processes nor with the general properties of jets and their relation to accretion disks.
These topics are reviewed and studied in many other papers where more references can be found (e.g., \citealt{Mirabel1999, Fender2004, Ferreiraetal2006, Livio2011, FenderBelloni2012, Pudritzetal2012, McKinneyetal2013, ZanniFerreira2013, Franketal2014, Lasotaetal2014, Lovelaceetal2014, McKinneyetal2014}).
I assume that whenever the accreted gas has a large enough specific angular momentum jets are launched.

The goal of this review is to emphasize the similarities in the ingredients and properties of the JFM that operates in very different types of objects.
These similarities allow us to learn from one type of objects on others. The study is concentrated on the similar processes of jet-medium interaction via a feedback cycle in the different types of objects. For that, many studies of the different objects and feedback processes in individual types of objects that do not consider the similarities between the different objects, will not be mentioned here.

The structure of this review is as follows.
In section \ref{sec:properties} I review some properties of the JFM that are common to many, or even to all, of the different types of systems reviewed here. Only a small number of references are given in that section.
In the sections to follow I review each class of systems separately, and implement the relevant ingredients of the JFM to the different classes.
Table \ref{table:compare} lists the types of objects that will be reviewed here, with some relevant properties.
The caption of the table defines different abbreviations and acronyms.
The properties that are listed in Table \ref{table:compare} will be explained and discussed along the manuscript, so the table should be used as a guide while reading this review.
\begin{table*}
\tiny
    \begin{tabular}{|l||l|l|l|l|l|l|l|l|}
     \hline
Property &Clusters & Galaxy-F & CCSNe & PNe & CEE&GEE&ILOTs&YSOs  \\
 \hline
  \hline
Energy (erg)&$10^{60}$&$10^{59}$&$10^{51}$&$10^{44}$&$10^{44-48}$& $10^{44-48}$ & $10^{46-49}$ &       $10^{43-46}$\\
             \hline
Mass $(M_\odot)$& $10^{12}$ & $10^{11}$  & $10$  &$1$  & $1$  & $1$ & $1-10$ & $1-10^3$\\
              \hline
Size; $R_{\rm res}$&$100 \kpc$&$10 \kpc$&$10^9 \cm$&$0.1 \pc$& $10^{1-2}R_\odot$& $10^2R_\odot$ & $10^2R_\odot$ & $10^{3-5} \AU$\\
              \hline
$-\Phi_{\rm res}$ &$(1)^1$&$(0.3)^2$&$(10)^2$&$(0.03)^2$ &$(0.03)^2$ & $(0.03)^2$&$(0.1)^2$ & $(0.001)^2$ \\
          \hline
Time  & $10^{7-8} \yr$ & $10^{7-8} \yr$ & $1-3 \s$ & $10^{1-2} \yr$ & $1-100\yr$ & $10-100\yr$ & $0.1-10\yr$ & $10^{2-5}\yr$  \\
              \hline
$T_{\rm bubble}$(K) & $10^{9-10}$ & $10^{9-10}$ & $10^{10}$ & $10^6 $ & $10^{7}$  & $10^{7}$ & $10^{7}$ & $10^{3}$\\
 \hline
$T_{\rm ambiant}$(K)& $10^{7-8}$ & $10^{6-7}$ &few$\times 10^9$ & $10^4$ & $10^{5-6}$& $10^{5}$ & $10^4$ & $100$\\
      \hline
ComObj & SMBH & SMBH & NS/BH  & MS/WD       & MS/WD/NS & MS & MS & MS     \\
 mass $(M_\odot)$ & $10^{8-10}$& $10^{6-9}$ & $1-50$ & $1$ & $1$ & $1$ & $1-10$ & $1-10$ \\
$R_a$(cm) & $10^{13-16}$ & $10^{11-14}$ & $10^{6}/10^{7}$ & $10^{11}/10^{9}$  & $10^{11}/10^{9}/10^{6}$ & $10^{11}$ & $10^{11-12}$ & $10^{11-12}$ \\

$-\Phi_a$ & $c^2$ & $c^2$ & $(0.1-1)c^2$ & $(0.5)^2/(5)^2$  & $(0.5)^2/(5)^2/$  & $(0.5)^2$ & $(1)^2$& $(0.5)^2$ \\
& &  & & & $(100)^2$  &  & & \\
\hline
 $\Phi_a/\Phi_{\rm res}$&$10^5$&$10^6$&100&100&100&100&3-100&$10^5$\\
\hline
Jets' main     &Heating&Expelling&Exploding&Shaping&Removing   &Removing&Reducing &Expelling\\
$\qquad$ effect&the ICM&     gas & the star& the PN&part of the&envelope&accretion&gas;\\
               &       &         &         &       &envelope   &        &         &Turbulence\\
    \hline
Role of & Maintain   &$M_{\rm BH}-\sigma$&Explosion &Not&Might     &Ensures &Not &Slowing  \\
the JFM & ICM        &correlation  &energy $\approx$&much&limit    &outer   &Much&star  \\
        & temperature&             &binding         &    &accretion&envelope&    &formation \\
        &            &             & energy         &    &rate     &removal &    &\\
    \hline
Observation&X-ray   &Massive&(Axi-)   &Bipolar& ~      &Bipolar&Radiation;&Bipolar\\
           &bubbles;&outflow&symmetry)&PNs    & ~      &remnant&bipolar   & outflows\\
           &cold gas&       &         &       & ~      &       &remnant   &  \\
    \hline
Fizzle &Cooling    & Rapid & BH        & &Core-    & Forming & &More           \\
outcome&catastrophe& SMBH+stars& formation;& &secondary& a common& &gas forms  \\
       &           & growth& GRB       & &merger   & envelope& &stars  \\
    \hline
Importance&Crucial in&Very &Contestant &Jets common;&Might occur &Crucial&Not    &Common;  \\
of jets/JFM&all CFs   &common; &with neutrino&the JFM does&in some cases;&&crucial&not   \\
          &          &not crucial&mechanisms&not operate &not crucial   &&       &crucial    \\
    \hline
Status of  &In       &In       &In fierce&In       &Not in   &Newly   &Not in   &In           \\
jets/JFM in&consensus&consensus&debate   &consensus&consensus&proposed&consensus&consensus\\
community  &         &         &         &         &         &        &         &           \\
\hline
    \end{tabular}
    \caption{Systems discussed in this paper where feedback and/or shaping by jets take place.
    The different listed values are typical and to an order or magnitude (or two even) accuracy only.
    Typical energy: Energy in one jet episode.
    Typical Mass, Size: of the relevant ambient gas. Typical time: the duration of the jets activity episode.
    In the row of observations, in parenthesis are expected observations.
    \newline
    Abbreviations and acronyms:
$R_a$ is the typical radius of the accreting object, and $\Phi_a$ is the magnitude of the gravitational potential on its surface in terms of the light speed $c$ or in units of $(1000 \km \s^{-1})^2$.
$R_{\rm res}$ stands for the typical radius of the reservoir of gas for accretion onto the compact object (the size of the system), and $\Phi_{\rm res}$ is the specific energy required to expel the reservoir (energy per unit mass) from the system in units of $(1000 \km \s^{-1})^2$.
  \textbf{Galaxy-F}: galaxy formation; \textbf{PNe}: Planetary nebulae; \textbf{CCSNe}: core collapse supernovae; \textbf{CF}: cooling flow; \textbf{ICM}: Intra-cluster medium; \textbf{CEE}: common envelope evolution; \textbf{GEE}: grazing envelope evolution;     \textbf{ILOTs}: intermediate-luminosity optical transients; \textbf{YSOs}: young stellar objects; \textbf{BH}: black hole; \textbf{SMBH}: super-massive BH; \textbf{NS}: neutron star; \textbf{WD}: white dwarf; \textbf{MS}: Main sequence star; \textbf{ComObj}: The compact object that accretes mass and launches the jets; $M_{\rm BH}-\sigma$ stands for the correlation of the SMBH mass with the stellar velocity dispersion. Based on \cite{Sokeretal2013}, \cite{Soker2015EAS} and \cite{KashiSoker2016}.
   }
 \label{table:compare}
\end{table*}

The systems to be studied are as follows. Planetary nebulae (PNe) share properties with all the other types of objects. For that, despite that the JFM does not act directly in PNe, they are reviewed here in section \ref{sec:PNe}. The X-ray deficient cavities, that are all jet-inflated bubbles, in cooling flows in groups and clusters of galaxies are the best observed entities that testify at the operation of the JFM. I review cooling flows in section \ref{sec:CFs}. In section \ref{sec:galaxy} I review the JFM that might have operated during galaxy formation, and point to possible similarities with the JFM that operates in cooling flows.
Then I move to the speculative and controversial scenario that jets explode core collapse supernovae (CCSNe), and that they operate through a JFM (section \ref{sec:SNe}).
In the common envelope evolution (CEE) the JFM is not fundamentally necessary to explain the observations though may be occurring. Nonetheless, in section \ref{sec:CEE} I suggest that in many cases jets facilitate the removal of the common envelope (CE), and that the jets operate through a JFM.
When the jets are very efficient in removing the envelope, a CEE might be prevented. Instead a grazing envelope evolution (GEE) takes place. An efficient JFM is necessary for the GEE to take place (section \ref{sec:GEE}).
The recent proposal that the JFM might operate in some of the transient objects termed intermediate luminosity optical transients (ILOTs: Red Novae, Red Transients, intermediate-luminous red transients, SN impostors, LBV variables) is reviewed in section \ref{sec:ILOTs}. In those systems the JFM is not necessary, and its operation may have low efficiency.
In section \ref{sec:YSOs} I conclude with the JFM that operates in some young stellar objects (YSOs). It has some similarity with the JFM in other type of objects, but also some significant differences. In particular, the JFM there operates in clumps (clouds), and several stars can launch jets into the same clump.
 I summarize this review with a list of open questions in section \ref{sec:summary}.

\textbf{$\bigstar$ Summary of section \ref{sec:intro}.} Table 1 summarizes the types of objects that I review in the coming sections. Despite the huge differences in energy, mass, size, jets' velocity, and typical temperatures, between the eight types of objects, there are many common properties in the accretion process and the interaction of jets with the ambient gas.

\section{GENERAL PROPERTIES OF THE JFM}
\label{sec:properties}

In this section I review several of the general processes and properties of the JFM.
I will further discuss these in the respective sections where I discuss the different systems, and where I list more references.

\subsection{The importance of jets}
\label{subsec:jets}

We can understand the importance of jets in delivering energy to the ambient gas from three simple and basic expressions. Those are the expression for the gravitational energy, the hydrostatic equilibrium equation, and the expression for the optical depth. Although they are well-known, I will nonetheless write them here to emphasize the important role of jets.

The first is the expression for the gravitational energy. The gravitational energy $E_a$ that is released by the accretion of mass $m_a$ on to the compact object of radius $R_a$ and mass $M_a$ is
\begin{equation}
E_a = - \Phi_a m_a = \frac {G M_a}{R_a} m_a,
\label{eq:phia}
\end{equation}
where the second equality defines the potential $\Phi_a$ on the surface of the compact accreting object.
The ambient gas in the system is the reservoir from which the compact object accretes mass. The binding energy of the gas reservoir in the system is
\begin{equation}
E_{\rm res} \simeq - \frac{1}{2} \Phi_{\rm res} M_{\rm res} \approx \frac {G M^2_{\rm res}}{ R_{\rm res}},
\label{eq:phir}
\end{equation}
where $\Phi_{\rm res}$ is the average potential of the ambient gas, $R_{\rm res}$ is the size of the reservoir, and $M_{\rm res}$ is the mass. The expression for the potential in the second equality in clusters and galaxies is not accurate, as for dark matter the potential well is deeper even.

The systems discussed here are much larger than the accreting compact object (beside in some ILOTs to be discussed later), $R_{\rm res} \gg R_a$, such that $\vert \Phi_a \vert \gg \vert \Phi_{\rm res} \vert$, as can be seen in Table \ref{table:compare}. Hence, a small amount of mass that is accreted by the compact object can significantly heat, expel, and inflate the ambient gas.

In principle radiation, neutrinos in core collapse supernovae (CCSNe) and photons in the other systems, can transfer the energy from the accreted gas to the ambient gas. However, in most cases this process is not efficient.  The reason is that the systems are in, or close to, hydrostatic equilibrium, either along the entire feedback cycle, or just before its operation. As the temperature gradient is never very steep, hydrostatic equilibrium implies that in most cases the density decreases quite rapidly with radius,
\begin{equation}
-\frac{1}{\rho(r)} \frac{dP}{dr} = \frac{d}{dr} \Phi_{\rm res} (r) \quad \rightarrow  \quad
\frac {d \ln \rho} {d \ln r} < -1 .
\label{eq:static}
\end{equation}
This in turn implies that the optical depth from radius $r$ to the edge of the system decreases rapidly with increasing radius. If we take the density to be $\rho \propto r^{-\alpha}$ with $\alpha >1$, and a constant opacity $\kappa$, the optical depth to infinity in most systems studied here is
\begin{equation}
\tau (r) = \int_r^\infty \rho \kappa dR \propto r^{1-\alpha} \approx {\rm steeply ~decreasing}.
\label{eq:tau}
\end{equation}
This implies that up to some radius the photons are trapped and diffuse slowly outward, then, from close to the photosphere they escape and do not deposit much energy.
An argument along this line, but limited to active galactic nuclei (AGN), is given by \cite{KingPounds2014}.
A very similar argument holds for energy carried out by neutrinos in CCSNe.

Although the argument presented above is very basic and simple, it has far reaching implications. For example, it can be used to show the severe limit on the energy that neutrinos can transfer to the explosion of massive stars (\citealt{Papishetal2015}; see section \ref{sec:SNe}). Jets instead, as will be discussed throughout this review, can efficiently interact with the ambient gas at all radii.

\subsection{Some basic ingredients}
\label{subsec:ingredients}

There are some common processes and properties of the JFM that are common to all, or many of, the different astrophysical objects studied here.
I list them here, and expand on different aspects in following sections, and implement them to the specific types of objects in the respective sections.

(1) \emph{Accretion disk and belt.} The jets are launched by an accretion disk around a compact object (e.g., \citealt{Livio1999}). In the case of jets launched by SMBH such disks are inferred from observations, and the jets are directly observed. In some other cases, such as in CCSNe if they are driven by jets as claimed in this paper, the jets cannot be directly observed.
{{{{ However, they can be indirectly inferred, such as by spectropolarimetric observations that indicate that some CCSNe are likely to have elongated morphologies \citep{Inserraetal2016}. }}}} The same holds for galaxy formation, as young galaxies are at large distances and in many cases the central region is obscured by dust.
In CCSNe (section \ref{sec:SNe}) the accreted gas might have a stochastic angular momentum, and the jets'  direction changes over time in a stochastic manner.
These are termed \emph{jittering jets}.
More over, in some cases, such as in the CEE with a main sequence companion and in some CCSNe, the accreted mater might have a specific angular momentum lower than that required to form an accretion disk around the accreting object. It is possible that in these sub-Keplerian accretion flows an accretion belt is formed around the accreting body, and that the belt launches jets \citep{SchreierSoker2016}.

(2) \emph{Universal jets' properties.} The properties of jets launched by the compact objects have some universal average properties (e.g., \citealt{Livio1999}).
$(i)$ The velocity of the pre-shock jets' material is $v_j \simeq v_{\rm esc}$, where $v_{\rm esc}$ is the escape velocity from the compact object.
$(ii)$ The ratio between mass lose rate {{{{ (of gas with a velocity about equals to the escape velocity) }}}} in the two jets to mass accretion rate is $\dot M_j/\dot M_{\rm acc} \approx 0.1$.

(3) \emph{Non-penetrating jets.} For a jet to efficiently deposit energy in the relevant inner regions it should not penetrate through these regions. The non-penetration condition is that the time required for the jet to propagate through the ambient gas and to exit from it, must be longer than the jet's effective activity time along a specific direction.
The effective activity time might be determined by the supply of gas to the jet, or by a relative transverse motion of the jets and the ambient gas. In the later case the time is the width of the jet divided by the relative transverse motion.

(4) \emph{Available mass.} The mass available for the inflow is very large. Namely, the mass that is eventually accreted to the compact object(s) is limited by the JFM and not by the mass available in the surroundings.

(5) \emph{Stochastic accretion.}
In some, but not all, cases, like in cooling flows, galaxy formation, and CCSNe, the feeding of the accretion disk might be in dense clumps.
Although the average mass accretion rate is regulated by the JFM, there might be large temporary variations in the  mass and angular momentum accretion rates.

(6) \emph{Incomplete energy deposition.} In closing a full negative feedback, the JFM does not influence the entire ambient gas with full efficiency.
For example, in cooling flows clusters and galaxies the accretion flow from the ambient gas is of cold clumps (up to the region where the accretion disk is formed). This implies that the jets, although very efficient, do not heat the entire intra-cluster medium (ICM) as to prevent the formation of cold clumps.
In the jittering jets model of CCSNe, as another example, the accretion continues for a few seconds while the jets and the bubbles they inflate expel most of the core mass. Namely, the first launching episode of the jets does not expel the entire core.

\subsection{The efficiency of the JFM}
\label{subsec:efficient}

The principle operation of the JFM is made possible by the deep potential well of the accreting object. There are two other processes that determine the efficiency of JFM.

\subsubsection {The non-penetrating condition}
\label{subsubsec:penetrating}

To efficiently interact with the ambient gas the jets must not drill a hole and escape out from the system. Narrow and dense jets that maintain a constant direction expand to large distances, and do not  deposit their energy and/or momentum in the inner regions. Such is the case with narrow AGN radio jets that reach huge distances in galaxies, but do not deposit much energy to the ISM. To prevent such a penetration to large distances, the jets and the ambient gas should obey one or more of the following conditions.

\noindent (1) \emph{Wide jets.} If the jets start wide, they interact with a large volume of the ambient gas, and slow down substantially. There is no unique value to define a wide jets but generally we can take it to be jets with a half opening angle of $\alpha_j \ga 30^\circ$.
In section \ref{sec:CFs} I will list several evidences to the presence of slow-massive-wide (SMW) outflows (jets) from AGN. It is quite possible that wide jets, that might be as well termed disk-winds, are very common in many types of objects where they are not observed directly.
{{{{ (I will hold to the term jet, rather than a wind, even when the bipolar outflow is wide and slow, because the general process of launching and the manner by which the bipolar outflow affects the ambient gas is the same as of jittering or precessing jets.) }}}}
\newline
(2) \emph{Subsonic jets.} This case becomes similar to the case of wide jets. A subsonic jet with a Mach number of $\mathcal{M}=v_j/c_s<1$, where $v_j$ is the jet velocity and $c_s$ is the sound speed inside the jet, has a thermal pressure larger than its ram pressure $\rho v^2_j$. Hence, the expansion of the jets to the sides will be significant. After a short travel distance the jet will have a large opening angle. Numerically subsonic jets and wide jets are inserted differently. But their behavior is  similar.
\newline
(3) \emph{Jet-ambient gas relative motion.} In these cases there is a relative motion between the jet and the ambient gas in a direction perpendicular to the jet propagation direction. The jets in these cases can be very narrow and dense. Although they penetrate very rapidly through the ambient gas, because of the change of the direction this propagation is not for a long distance, and the energy and momentum are deposited in the inner region. I list four causes of transverse relative motion.
\newline
(3.1) Precessing jets. Precession is likely to be caused by a binary companion. This can be the case in stellar systems, and also in AGN where there are binary systems of super-massive black holes (SMBHs). In PNe such jets are thought to be common.
\newline
(3.2) Relative motion of the source. When the jets' source is moving relative to the ambient gas, the jets will have a relative motion to the ambient gas as well. Such is the case when the jets' source is a star orbiting its companion, as in the CEE, the GEE, and ILOTs, or a SMBH that has a relative motion to the ISM or ICM.
\newline
(3.3) Ambient turbulence. Although there is no global relative motion between the jet and the ambient gas, the turbulence in the ambient gas implies that the jet has to encounter fresh gas all the time. Namely, ambient gas streams to the hole that the jet has drilled. When the turbulence is strong, e.g., as in the envelope of asymptotic giant branch (AGB) stars, this process might substantially slows down the jets, and their energy is deposited in the inner region.
\newline
(3.4) Jittering jets. When the accreted gas has a varying angular momentum axis, the launching direction of the jets varies as well, possibly in a stochastic manner. This might be the cases in some AGN, and in the explosion by jets of some CCSNe.
{{{{ \cite{PettiboneBlackman2009} calculated how inhomogeneities in the accretion disk lead to angular momentum fluctuations and then to stochastic wobbling (jittering) of the jets. They further discuss such jets in blazars, YSOs, and binary stellar systems of pre-PNe and microquasars. Jittering (wobbling) jets might be a common phenomenon.  }}}}

\subsubsection {Radiative losses}
\label{subsubsec:losses}

As the jets interact with the ambient gas they are shocked. The jets' gas is shocked to very high temperatures, and loses energy by radiation. If the radiative cooling time of the shocked gas, $t_{\rm cool}$,  is longer than the flow time $t_f$, the energy of the jets is deposited to the ambient gas. The energy can be transferred to kinetic energy of the ambient gas, either ordered motion or turbulence, and/or it can heat the ambient gas. This type of interaction is termed energy-conserving interaction.
If, on the other hand, the radiative cooling time is shorter than the flow time, most of the thermal energy of the shocked jets' material is carried away by radiation. The radiation carries a relatively small amount of momentum. In this case the interaction is termed momentum-conserving.
In the momentum-conserving interaction the jets are much less efficient in influencing the ambient gas.
The JFM operates in the momentum-conserving case mainly in young stellar objects (YSO). In most of the other types of objects the inequality $t_{\rm cool} \gg t_f$ holds, and the energy-conserving case prevails.

We can further sort the energy conserving case into two subcases. In the first the medium is optically thin, $\tau \ll 1$, and the emissivity
$\dot e_V$, which is the radiative power per unit volume, is low. The radiative cooling time is then giving by
\begin{equation}
t_{\rm cool} \simeq t_{\rm rad} \equiv \frac{e_V}{\dot e_V} = \frac{5}{2} \frac {nkT}{\Lambda n_e n_p} \quad {\rm for} \quad \tau \ll 1,
\label{eq:taurad}
\end{equation}
where $n$, $n_e$, and $n_p$ are the total, electron, and proton number density, and $\Lambda (T)$ is the cooling function which depends on temperature and composition.
In this equation the emissivity is taken for photons emitted by an ionized gas
$\dot e_V=\Lambda n_e n_p$. In case of neutrino cooling in CCSNe a different equation should be used. The coefficient $5/2$ in the expression for the energy per unit volume $e_V=(5/2) nkT$ is taken for a case that the gas cools in a constant pressure.
The flow time is the typical size or radius of the system where interaction takes place, divided by the jet's speed. This subcase is common in cooling flows, during galaxy formation, and in PNe.

In the other energy-conserving subcase the cooling time is long because the photon diffusion time is longer than the flow time. Such is the case inside stars, i.e, in CCSNe, in the CEE, in the GEE, and in some ILOTs. The diffusion time of photons in the radial direction through a shell of mass $M_s$ and width $\Delta r \approx r_s$, is given by
\begin{equation}
t_{\rm cool} \simeq t_{\rm diff} \approx 3 \tau \frac{r}{c} \approx \frac {M_s \kappa} {4 c r_s} \quad {\rm for} \quad \tau \gg 1,
\label{eq:taudiff}
\end{equation}
where $c$ is the speed of light. It can be seen that this case occurs for a very high ambient density.

\subsection{Morphology}
\label{subsec:morphology}

If radiative losses are small, then the post shock jet's material is hot, and it expands and forms a low-density bubble inside the ambient gas.
As usually there are two opposite jets, a structure of two opposite bubbles is formed.
A pair of bubbles (or several pairs) is a generic structure of the energy-conserving JFM,
when the bubbles can be observed.
This structure is termed bipolar in the case of PNe.
In most YSOs, on the other hand, the jets interact in the momentum-conserving case, and no large bubbles are observed.
The most prominent structures are observed in cooling flows (clusters, groups, and galaxies) and in PNe. They are much better resolved in the visible band in PNe than in X-ray band for cooling flows. This is the main reason for discussing PNe in this review.

\begin{figure}
\centering
\hskip -1.5 cm
\includegraphics[width =80mm]{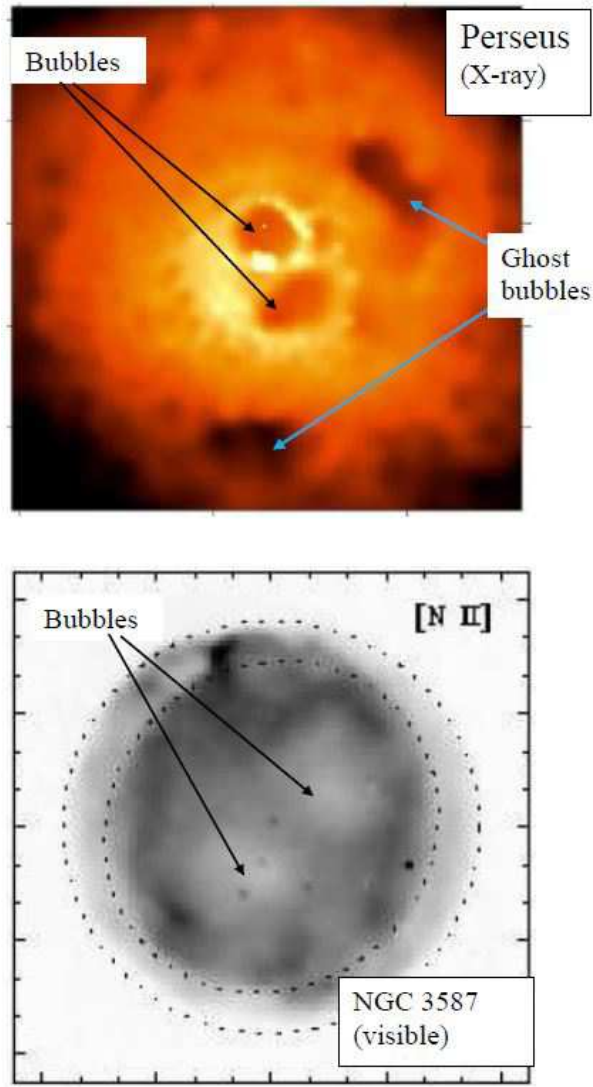} 
\caption{Comparing the morphologies of bubbles in a cluster  and in a planetary nebula. Presented are a false color Chandra X-ray image of the Perseus cooling flow cluster of galaxies (\citealt{Fabianetal2000}) and a visible image of the Owl (NGC~3587) planetary nebula \citep{Guerreroetal2003}.
Both objects display a clear pair of `fat' spherical bubbles near the center. In Perseus we can as well see a pair of bubbles that was inflated in a previous cycle of AGN activity. These are termed ghost bubbles.
 }
\label{fig:bubbles1}
\end{figure}

\begin{figure}
\centering
\hskip -1.5 cm
\includegraphics[width =80mm]{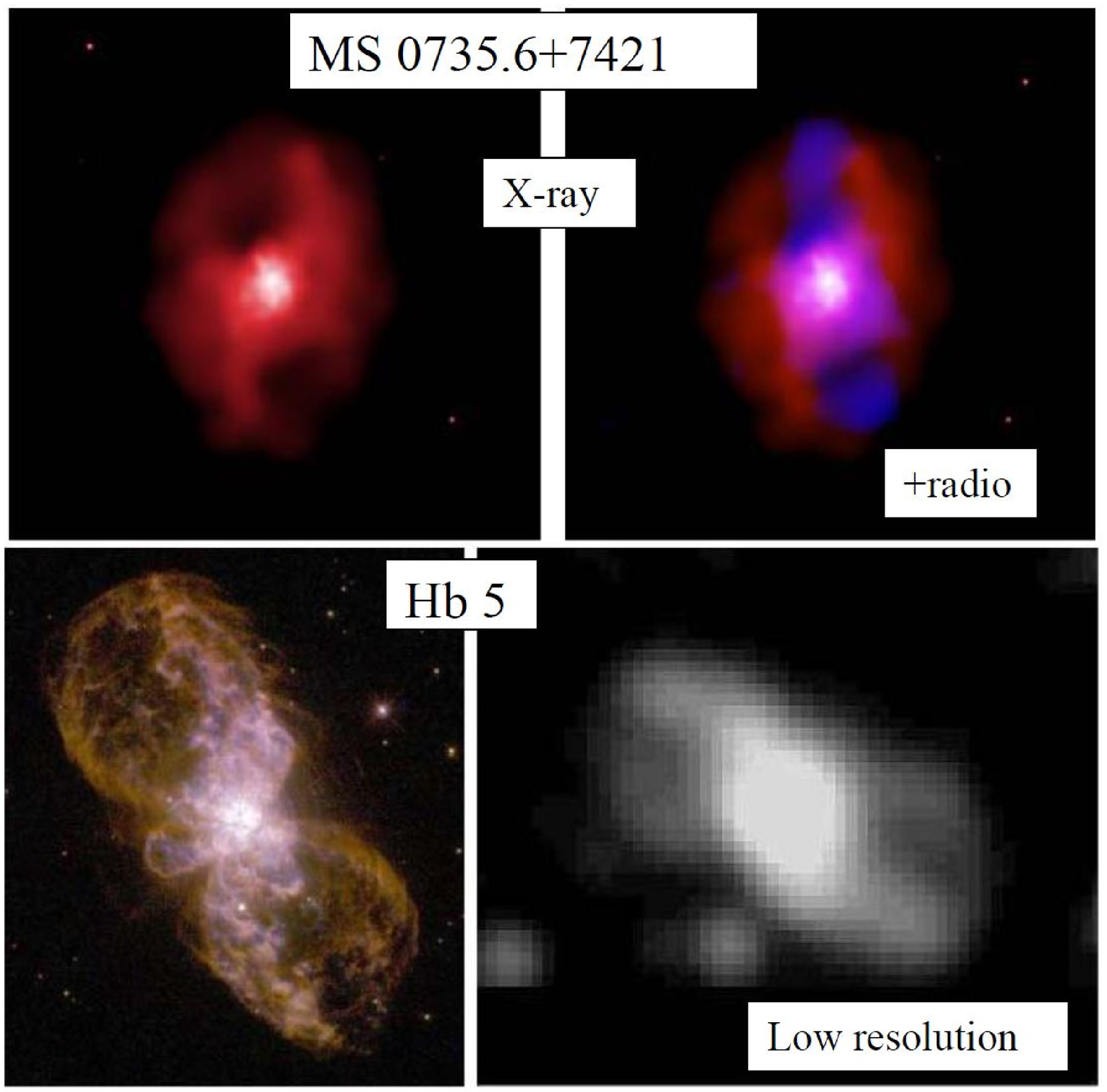} 
\caption{Like Figure \ref{fig:bubbles1} but emphasizing a point-symmetric morphology of the bubble pair, and the consequences of a bad resolution (right image of the planetary nebula; \citealt{Schwarzetal1992}). The right top panel display the radio emission inside the X-ray deficient bubbles.
The Chandra X-ray image of the MS~0735.6+7421 cluster is based on \cite{McNamaraetal2005}, and the left high-resolution Hubble Space Telescope image of the Hubble~5 planetary nebula is from \cite{TerzianHajian2000}.  }
\label{fig:bubbles2}
\end{figure}

There are many similarities in the morphologies of X-ray deficient bubbles observed in cooling flows and the bipolar structure of PNe. Bubble pairs in cooling flow clusters appeared already in results from the ROSAT X-ray satellite (e.g., \citealt{Boehringeretal1993, HuangSarazin1998}), {{{{ and brought the understanding that jet-inflated bubbles can heat the gas (e.g., \citealt{Churazovetal2000} for the \astrobj{Perseus} cluster and \citealt{Churazovetal2001} for the \astrobj{Virgo} cluster). }}}} But only the higher resolution images obtained by the Chandra X-ray Telescope boosted the study of the JFM in cooling flows (section \ref{sec:CFs}).
The bipolar structure of some PNe was known for a long time, but it took the community a decade to recognize that most of the bipolar bubble structures as well as other features are shaped by jets (section \ref{sec:PNe}).
Here I only present some similarities in the bubble morphologies of cooling flows and PNe (for more details and more examples see the arXiv version of \citealt{SokerBisker2006}).
The comparisons are presented in Figure \ref{fig:bubbles1} and \ref{fig:bubbles2}.

In Figure \ref{fig:bubbles1} the structure of a pair of almost spherical bubbles that touch the center is presented in the cooling flow cluster Perseus and in the Owl PN.
Despite the several orders of magnitude differences in size, energy, mass, and timescales, the similarity is not only in the morphology, but might be in some basic physical processes as well
\citep{Soker2003PASP, SokerBisker2006}.
I note that \cite{Franketal1993} suggested a different explanation for the bipolar structure of the Owl PN. In their explanation the slow AGB wind was dense near the equatorial plane and formed a torus. A spherical-fast wind is then turned on as the star leaves the AGB and forms a hollowed region elongated in the polar directions. When viewed at an angle, two low-intensity bubbles appear. Projection effects definitely play a role, even if jets shaped the bubbles, and can enhance the contrast. As well, enhanced equatorial mass loss as in the model of \cite{Franketal1993} can take place before, during, or after, the jets-launching episode (if jets exist). Here I adopt the view that jets played a role in shaping the Owl PN, but further work is required to find the exact shaping mechanism of the Owl and similar PNe.

Figure \ref{fig:bubbles2} emphasizes the point-symmetric structure of the MS~0735.6+7421 cluster, which hosts the most energetic radio active galactic nucleus (AGN) known \citep{Vantyghemetal2014}, and compares it with the visible image of the Hubble~5 PN.
The X-ray image of MS~0735.6+7421 and the low-resolution visible image of Hb~5 are very similar. The high resolution image of HB~5 hints at what we might be missing by lacking a high resolution X-ray image of MS~0735.6+7421.
Point symmetric PNe like Hb~5 are thought to be shaped by stellar binary interactions that cause a jet-precession. The similarity of the morphology of Hb-5 and the pair of bubbles in the CF cluster MS~0735.6+7421 brought \cite{Pizzolato2005b} to suggest that MS~0735.6+7421 has a SMBH binary system in its center.

\subsection{When the JFM fizzle}
 \label{subsec:fizzle}

I discuss here the outcomes when the JFM fizzle, i.e., operates at a very low efficiency.
This might be the case if the non-penetration condition is violated, as is expected if the jets
are well collimated and preserve a non-variable axis, and/or if radiative cooling is too rapid. The result of the JFM fizzle is listed in Table \ref{table:compare}.
It should be emphasized here and throughout the review that for some systems, in particular the CEE, the GEE, and the jittering-jets model of CCSNe, the JFM is still a speculative scenario that is not in the consensus. In these cases the fizzle-outcomes is speculative as well.
\newline
 \textbullet \emph{Clusters CFs and young galaxies.}
When the AGN jets do not heat the ICM efficiently, a cooling catastrophe might occur. The ICM (or ISM) cools on its radiative cooling time, and leads to high rate of star formation. Mass feeds the SMBH as well, and eventually energetic jets might heat the ICM to reestablish the JFM.
\newline
 \textbullet \emph{CCSNe.} The JFM mechanism can be inefficient in CCSNe when
the pre-collapse core is rapidly rotating. In this case a well defined accretion disk is formed around the newly formed NS, and the jets have a well define propagation axis. The jets do not
eject much of the core gas near the equatorial plane. A BH might form, and launch relativistic jets. This is the scenario for super-energetic SNe (SESNe) and gamma ray bursts (GRB) in the jittering-jets model. Eventually the jets do remove large portion of the stellar mass, and a CCSN does take place in parallel to the GRB.
\newline
 \textbullet \emph{CEE.} Jets might not be necessary to eject the envelope in
all CEE cases. In many cases where jets are not efficient, however, I speculate that merger of the secondary star with the core takes place. This might especially be the case with WD
companions spiraling inside the envelope of a relatively massive giant, more than about $3 M_\odot$. The merger product might be a Type Ia supernova progenitor according to the core-degenerate scenario.
\newline
 \textbullet \emph{GEE.} The GEE is based on efficient envelope gas removal by
jets. This prevents the formation of a CE. When the jets fail in
that, a CEE will commence.

\textbf{$\bigstar$ Summary of section \ref{sec:properties}.} This section can be summarized as follows. There are some basic properties of jets and the JFM that are shared by many or all of the types of systems listed in Table 1. When observed, bipolar bubbles is a generic structure of the interaction of jets with the ambient gas when radiative cooling is not significant (not in most YSOs). As will be discussed in the following sections, these are the bubbles that have the largest effect in shaping and energizing the ambient gas. In many of the cases there are processes that prevent the jets from expanding out of the ambient gas, by that enhancing the efficiency of the JFM. These can be ordered processes, as wide jets, as well as processes that break the generic axial-symmetry of the jet-launching process. The later include precessing jets, jittering jets, turbulence in the medium, and a motion (such as orbital motion) of the compact object that launches the jets relative to the ambient gas.  Some energetic and exotic phenomena can take place when the JFM efficiency substantially drops (fizzle).

\section{Planetary Nebulae (PNe)}
\label{sec:PNe}

In PNe gravity is negligible and there is no need for the JFM to operate. The JFM hence
does not act directly in the shaping of PNe.
Nonetheless, in some PNe and other nebulae around stars most of the kinetic energy of the expanding gas comes from the jets (e.g., \citealt{Bujarrabaletal2001}).

The reason I discuss PNe is that PNe are sitting at the cross-road of many other astrophysical objects.
For the present study most significant are that the morphologies of PNe are well resolved (e.g., \citealt{Balick1987, Chuetal1987, CorradiSchwarz1995, Manchadoetal1996, SahaiTrauger1998, Parkeretal2006, Parkeretal2016}) and share many properties with many other objects, and that the jets are launched mainly by a stellar companion that accretes mass from the giant progenitor.
The bipolar morphologies are related to the study of cooling flows and galaxy formation, and the binary interaction is relevant mainly to the CEE, the GEE, and possibly to YSOs and ILOTs.

\subsection{PN morphologies}
\label{subsec:PNmorphologies}

The morphologies of some PNe are observed to be similar to X-ray deficient bubbles in cooling flows \citep{SokerBisker2006}, as presented in Figures \ref{fig:bubbles1} and \ref{fig:bubbles2}, and to some nebulae around YSOs (e.g. \citealt{Leeetal2003, LeeSahai2004}).
Examples include the well collimated jets from Hen~2-90 that are composed from chain of bullets that are similar to jets from some YSOs \citep{SahaiNyman2000}, the similarity of the pre-PN CRL618 to YSOs \citep{Leeetal2003, SokerMcley2013}, and the striking similar morphology of the PN KjPn~8 (for a study of this PN see, e.g., \citealt{BoumisMeaburn2013}) and the bipolar structure of the recently discovered YSO Ou4  \citep{Acker2012, Corradietal2014}.

PNe were known to possess bipolar structures long before similar structures have been found in X-ray images of clusters of galaxies. The presence of jets in PNe was also deduced from observations
more than 30 years ago (e.g., \citealt{Feibelman1985}). \cite{Giesekingetal1985} found a collimated outflow in the PN NGC 2392. They mentioned the similarity of these jets with some jets in YSOs, and further speculated that such outflows exist in many PNe.
However, it took the community some time to realize the crucial role of jets in shaping many PNe (see review by \citealt{BalickFrank2002}).

On the theoretical side, \cite{Morris1987} suggested that two jets launched from the accretion disk around a stellar companion form a bipolar PN. The bipolar morphology of many symbiotic nebulae which are known to be shaped by jets (e.g., \citealt{Schwarzetal1989, CorradiSchwarz1995}) strengthen this shaping model of bipolar PNe.
The next step was the recognition that the two opposite clumps, called fast low-ionization emission regions (FLIERs or {\it ansae}) along the symmetry axis of many elliptical PNe are formed by jets, probably launched during the last phase of the AGB or the post-AGB phase of
the PN progenitor \citep{Soker1990AJ}.
Then \cite{SahaiTrauger1998} further argued, based on their high quality HST observations of PNe,
that the non-spherical structures of many PNe are formed solely by jets.

Many new observations and theoretical studies since the turn of the Millennium further consolidated the jet-shaping scenario of many PNe
(e.g., \citealt{Parthasarathyetal2000, Bujarrabaletal2001, Mirandaetal2001, Corradietal2001, Guerreroetal2001, Imaietal2002, Rieraetal2003, Vinkovicetal2004, Hugginsetal2004, Penaetal2004, BalickHajian2004, Arrietaetal2005, Oppenheimeretal2005, Sahaietal2005, Vlemmingsetal2006, Mirandaetal2010, Nakashimaetal2010, Phillipsetal2010, Sahaietal2011, Szyszkaetal2011, Lopezetal2012, Clarketal2013, Peresetal2013, Vlemmingsetal2014, Fangetal2015, Gomezetal2015, Akras2016, Clyneetal2015, Danehkar2015, Danehkaretal2016}, for observational papers, and, e.g.,
\citealt{SokerRappaport2000, Blackmanetal2001, LivioSoker2001, Soker2002, LeeSahai2003, GarciaArredondoFrank2004, Velazquezetal2004, Velazquezetal2007, Rieraetal2005, Ragaetal2008, StuteSahai2007, Akashietal2008, AkashiSoker2008MN, Dennisetal2008, Dennisetal2009, Leeetal2009, HuarteEspinosaetal2012, Velazquezetal2012, Balicketal2013, Velazquezetal2014} for theoretical studies; these papers composed a small fraction of the papers studying jets in PNe).
X-ray observations seem also to support the presence of jet-inflated bubbles and of jets
(\citealt{Kastneretal2003, Sahaietal2003}), similar to X-ray jets in symbiotic systems (e.g., \citealt{Kelloggetal2001, GallowaySokoloski2004}).

Again, the many well resolve PNe with bipolar bubbles have brought progress in the theoretical understanding of jet-inflated bubbles in PNe. People who study the JFM and the inflation of bubbles by jets in the other classes of systems studied here are encouraged to compare their results with those obtained by the PN community.
I consider as an example the shaping and flow in the process of inflating `fat' bubbles. Fat bubbles stand here for more or less spherical bubbles attached to their origin.

\cite{Sokeretal2013} compare  numerical simulations of jet-inflated bubbles conducted with conditions appropriate to PNe with those conducted for cooling flows. When the jets precess or have a wide opening angle, a half opening angle of $\alpha_j \ga 30^\circ$, there are some generic properties. The observational evidence for such wide jets will be reviewed in section \ref{sec:CFs}.
\newline
(1) The two opposite jets form a bipolar structure of two `fat bubbles', i.e., almost spherical bubbles almost touching the center.
\newline
(2) A dense thin shell is formed around the bubbles.
\newline
(3) The gas filling the bubbles, which is the shocked jets' gas, is much hotter than the shocked ambient gas. This does not necessary holds in bubbles inflated in the core of massive stars that are exploded by jets.
\newline
(4) Vortices are formed on the sides of the bubbles. The vortices are more pronounced in cooling flows.
\newline
The vortices to the sides of the bubbles might compress gas to a thin disk in the equatorial plane. This effect is more prominent in PNe.
\cite{AkashiSoker2008NewA} suggested that part of the equatorial dense gas might flow back to the center binary system of the PN and form a large (up to hundreds of AU) Keplerian disk. If some of this gas is accreted back to the center, we have an effect of a positive feedback.
\cite{Klassenetal2014} found a similar effect, of bubble pushing gas back to the accretion disk, in their simulation of bipolar bubbles inflated by radiation during the formation of massive stars.

Many of the well resolved PNe display a point symmetric morphology, namely, several symmetry axes rotate with respect to each other through a common origin.
The most likely explanation for the point symmetric morphology are precessing jets (e.g.,
\citealt{GuerreroManchado1998}) launched by a precessing accretion disk in a binary system.
This brings us to the second important property of PNe, that binary interaction is behind the formation of jets.

\subsection{Binary interaction}
\label{subsec:PNbinary}

The connection between the presence of jets (or collimated polar outflows) in PNe (and in pre-PNe) and the progenitor being in a binary system has received a huge support in the last decade (e.g., \citealt{Miszalskietal2009, Ragaetal2009, Wittetal2009, Corradietal2011aa, Corrdietal2011MN, Miszalskietal2011,  Velazquezetal2011ApJ, Boffinetal2012, Miszalskietal2013, Corradietal2014MN, Tocknelletal2014, Akarasetal2015, Chenetal2016}, to list a small fraction of the relevant studies). Not only white dwarf and main sequence stellar companions can shape PNe, but brown dwarfs and massive planets can play a non-negligible role as well
(e.g., \citealt{Soker1996, NordhausBlackman2006, DeMarcoSoker2011, Staffetal2016MN8}).

The fundamental role of the binary system is to supply the required angular momentum.
Evolved stars lose most of their angular momentum in strong winds as they become giants. To have an accretion disk or an accretion belt that are thought to launch jets, a source of angular momentum is required. The companion can spin up the giant star, or more likely for the formation of jets in PNe, accrete mass from the giant star. The accreted gas has enough angular momentum to form an accretion disk or an accretion belt around the companion.

Interesting effects result from a companion that launches jets as it orbits the primary star. If the orbit is eccentric, for example, the morphology of the descendant PN will possess departure from axi-symmetry. In another effect the AGB wind can deflect the jets launched by the companion. \cite{GarciaArredondoFrank2004} were the first to conduct 3D numerical simulations
of the interaction of jets launched by a secondary star with the slow primary wind. Their high quality results strengthen the general stellar-binary jets model, and the conjecture \citep{SokerRappaport2000} that a narrow waist can be formed by the jet.

In the JFM model for super-energetic supernovae (SESNe; section \ref{sec:SNe}), a binary companion is required to spin-up the core. The binary companion must therefore spiral-in inside the CCSN progenitor to reach the core. It will have another effect of removing the hydrogen-rich envelope, and possibly the helium-rich envelope. But what might be very interesting is if the companion survives to the time of the explosion with a small orbital separation, $< 1 R_\odot$. The newly born neutron star (NS) will launch jets as it orbit around the center of mass. The SN explosion might possess departure from axi-symmetry in a manner similar to some PNe.

Another support to the connection of strong binary interaction and the formation of bipolar bubbles comes from the Homunculus, the bipolar structure around the binary star $\eta$ Carinae. It is very likely that this bipolar structure was also shaped by bipolar jets \citep{Soker2001Car,Soker2005Car}. These jets could have been  launched by the companion as it accreted mass at periastron passages during the 1837-1856 Great Eruption (\citealt{Kashisoker2010Car}). Supporting this model is the finding by \cite{Staffetal2016MN} that a main sequence companion can accreted mass at a relatively high rate at periastron passages in its orbit around an AGB star.

\cite{BlackmanLucchini2014} conduct a simple and beautiful study that demonstrates the usage of jets in pre-PNe to infer about the CEE. They use the observed momenta of the bipolar outflows in 19 pre-PNe to constrain the viable accretion engines for bipolar outflows in the progenitor. Their main conclusion is that in most systems the accretion rate is super-Eddington, and it requires that the accretion process takes place inside a common envelope. I here add that the GEE also fits these results. It seems that main sequence stars in a CEE and GEE can indeed accrete mass at super-Eddington rates. In the GEE and CEE the jets activity might act in a negative feedback cycle.

\textbf{$\bigstar$ Summary of section \ref{sec:PNe}.} I summarize this section by listing the two insights that PNe give us. (1) Hot bubbles are a generic outcome when jets interact with ambient gas. The same is observed in X-ray deficient bubbles (cavities) in cooling flows. (2) Strongly interacting binary systems can launch energetic jets. This adds to the motivation to consider the operation of the JFM in the GEE, in some cases of the CEE, and in some ILOTs.

\section{Cooling Flows}
\label{sec:CFs}

Cooling flow is the name given to the processes where ICM gas in clusters (or groups) of galaxies,
or the ISM in elliptical galaxies, loses energy by radiation, mainly in the X-ray band, and part of the gas cools further and flows inward.
The cooling to low temperatures is evident from star formation (e.g., \citealt{Odeaetal2010}), molecular gas (e.g., \citealt{Wilmanetal2009}), and dust (e.g., \citealt{Edgeetal2010}). The mass cooling rate to low temperature is much lower than what would be the case if the ICM was not heated. Hence, the ICM must be heated, most likely in a negative feedback cycle.
Because of this lower cooling rate to low temperatures, some people have abandon the name cooling flow, and use cool core. In a previous paper \citep{Soker2010critical} I criticize this change of name, and point out that the PNe community did not change the name even after PNe have been shown not to be planets.

Cooling flows in clusters of galaxies and in galaxies are the only type of objects among those reviewed here where the JFM is observed in its full glory. We can determine both the jets' power and the cooling power, and observe consequences of different processes.
In cooling flows the energy source of the jets is an AGN, and the general feedback cycle is termed AGN feedback. The AGN can supply mechanical energy of jets and radiation energy. For the goals of the present review I concentrate on jets launched by the AGN, hence this might be termed AGN-JFM.

\cite{Fabian2012} and \cite{McNamaraNulsen2012} review many, but not all, aspects of the AGN feedback. An earlier review is given by \cite{McNamaraNulsen2007} where some more details of cooling flows are given.
I will not repeat topics reviewed by these authors unless they are relevant to the comparison with the other types of objects reviewed here. I will also mention some aspects of the AGN-JFM that were not mentioned in these two reviews, e.g., the role or vortices.

\subsection{Jet-inflated bubbles}
\label{subsec:CFbubbles}

The understanding that the ICM in cooling flows should be heated by some negative feedback mechanism goes back more than 20 years (e.g., \citealt{Binney1995, Churazovetal2002}). Over the years the AGN feedback mechanism, mostly by jet-inflated bubbles, has been solidified
(e.g., \citealt{Dong2010, OSullivan2011, Farage2012, Gaspari2012a, Gaspari2012b, Birzan2011, Gitti2012, Pfrommer2013}). Key to the establishment of the AGN-JFM in cooling flows are the many observations of X-ray deficient bubbles (e.g., \citealt{Boehringeretal1993, Fabianetal2000, McNamara2001, Heinz2002, Formanetal2007, Wiseetal2007, Baldi2009, David2009, Gastaldello2009, Gitti2010, Blanton2011, David2011,  Machacek2011, Randalletal2011, Doria2012, Pandge2012, Hlavaceketal2015}).
Many of these X-ray  bipolar structures resemble visible bipolar structures observed in bipolar PNe (see Figs. \ref{fig:bubbles1} and \ref{fig:bubbles2}). As I discuss in section \ref{subsec:CFheating}, the similarity goes beyond the morphology.

The energy that is needed to inflate a bubble with a volume $V$ inside the ICM of pressure $P$ is given by $E_{\rm bubble} = 4 PV$ for a bubble filled with relativistic gas, and by $E_{\rm bubble} = (5/2) PV$ for a non-relativistic filling. The power of the jets is crudely estimated by taking the inflation time to be about equal to the buoyancy rise time of the bubble (e.g., \citealt{Hlavaceketal2015}). Studies by, e.g.,  \cite{Churazovetal2000}, \cite{Churazovetal2001}, \cite{Raffertyetal2006} and \cite{Hlavaceketal2015}, show that the power of jets that inflate bubbles is about equal or somewhat larger than the radiative cooling of the ICM.
 This result serves as a very strong support to the AGB-JFM in cooling flows.

In clusters the ambient gas has a temperatures of $\approx 10^7 \K$, and the gas inside the bubble has a temperature of $T > 10^9 \K$ (not determined yet). In PNe, the temperature of the ambient gas is $10^4 \K$, and that of the gas inside the bubble is $\approx 10^6 - 10^7 \K$ (e.g., \citealt{Guerreroetal2002, Freemanetal2014}).
The great advantage of clusters is that the ICM are transparent to both X-ray, where the ambient gas is seen, and to radio that reveals the interior of the bubbles in many cases. In PNe the ambient region is always seen in visible, but only in rare cases do we see the inner region of bubbles because of the weak X-ray emission, e.g., in the PN \astrobj{NGC~6543} (the Cat's Eye Nebula, \citealt{Chuetal2001}). X-ray have been found in the PN \astrobj{Hb~5} that is presented in Figure \ref{fig:bubbles2} (\citealt{Montezetal2009, Freemanetal2014}), and in a handful of other PNe \citep{Freemanetal2014}. Although a small number, these cases of PNe are sufficient to demonstrate the similarity in the process of inflating bipolar bubbles with jets in PNe and in cooling flows
\citep{Soker2006aph}.

There are some properties of the bubbles in clusters cooling flows that cannot be revealed yet from observations. One is the temperature of the gas filling the bubble, or the general content of the interior of the bubble. {{{{ This is because of the very low density of the gas inside the bubbles in cooling flows, unlike in PNe, implying luminosity much below detection limits. It is possible that magnetic fields dominate the energy content of bubbles (e.g., \citealt{Kronbergetal2001}). }}}}
Another one is the structure of many small-scale vortices that develop inside the bubble and in the ICM in the vicinity of the bubble. Vortices play important roles in entrainment of gas by the jets and bubbles (e.g., \citealt{Ommaetal2004, GilkisSoker2012}), in the stability of bubbles (see below), and most important in the mixing of hot bubble gas with the ICM. This mixing is a most important process, as it might be the main heating process of the ICM (as discussed in section \ref{subsec:CFheating}).

An issue that is often discussed is the stability of bubbles (e.g., \citealt{McNamaraNulsen2012}  list processes that have been suggested to stabilize bubbles). Observations suggest that bubbles in cooling flows live for a long period of time (e.g., \citealt{Fabianetal2011}), but some development of Rayleigh-Taylor instabilities are seen at the center-front of some bubbles,
e.g., in \astrobj{Perseus} \citep{Fabianetal2002} and in \astrobj{A~2052} \citep{Blanton2011}.
The protrusion in the front of the northwest bubble in \astrobj{Perseus} seems to be compatible with the evolution time of a Rayleigh-Taylor instability \citep{Sokeretal2002}.

In the last decade more and more numerical simulations inflate bubbles with jets (e.g.,  \citealt{Sternbergetal2007, FalcetaGoncalvesetal2010, Morsonyetal2010, Gasparietal2011}, and papers since then), rather than inserting artificial bubbles by inflating a sphere of hot gas some distance form the center (as was used to be the case in the more distant past).
In such cases there is no need to invoke any particular process, such as magnetic fields {{{{ (even if they exist), }}}} to stabilize bubbles in cooling flows. (It is not clear at all how magnetic fields can stabilize bubbles, {{{{ as magnetic tension stabilizes along the filed lines, but not perpendicular to the field lines). }}}}  \cite{SternbergSoker2008b} demonstrate that the key to obtain the correct evolution of bubbles and their interaction with the ambient gas in numerical simulations is to properly inflate them by jets (see discussion in \citealt{Soker2010critical}). The results then are as follows.
(1) The deceleration of the bubble-ICM interface during the bubble-inflation phase stabilizes the bubble  (\citealt{Sokeretal2002, PizzolatoSoker2006}).
(2) The dense shell formed at the front of the bubble adds to the stability.
(3) Vortices inside the jet-inflated bubble stabilize the sides of the bubble as the bubble rises through the ICM \citep{SternbergSoker2008b}. Vortices play a key role in heating the ICM as well.

In contrast, when bubbles are inflated in a short time from a point rather by a jet the interface of the bubble with the ambient gas accelerates at early times, and instabilities develop on the interface. This is the case for example, when jets interact with the circum-stellar wind in a short time, as in simulating the formation of a bipolar PN in an ILTO event \citep{AkashiSoker2013}). However, these small scale instabilities do not destroy the large-scale bipolar morphology of PNe.

Some instabilities might act on the jet and its cocoon (its post-shock gas that trails behind). In some cooling flows two opposite chains of bubbles that are close to each other, and even overlap, are observed, as in \astrobj{Hydra~A} \citep{Wiseetal2007}, and in two bubbles in the galaxy group \astrobj{NGC~5813} \citep{Randalletal2011, Randalletal2015}.
These chains are usually attributed to several episodes of jet activity.
\cite{RefaelovichSoker2012} show in two-dimensional hydrodynamical simulations of a continuous jet that several fragmentation mechanisms that act on the primary vortex that is created just behind the jet's head, might split it to several smaller vortices. this chain might then appear as a chain of bubbles.

\textbf{$\bigstar$ Summary of subsection \ref{subsec:CFbubbles}.} Bubbles in cooling flows (termed also X-ray deficient bubbles) possess bipolar morphologies with some similarities to those in PNe, and some massive stars such as \astrobj{$\eta$~Cariane}. The energy content in the bubbles is sufficient to offset radiative cooling in cooling flows, suggesting they are part of a feedback cycle. Numerical simulations that inflate bubbles with appropriate jets (see next section) can explain the slow development of instabilities in the bubbles, with no need to introduce extra stabilizing effects.

\subsection{The Jets}
\label{subsec:CFjets}

Many bubbles are wide, almost spherical, and touch the center of the AGN that launches the jets. These are termed `fat bubbles'.
The conditions to inflate fat bubbles were derived in some numerical simulations, with the basic result that the jet should obey the non-penetrating condition (section \ref{subsubsec:penetrating}). This implies that one or more of the following should hold.
The jets are slow (sub-relativistic),  massive, and wide (SMW) jets, or bipolar outflows
(e.g., \citealt{Sternbergetal2007}), the jets precesses (e.g., \citealt{SternbergSoker2008a, FalcetaGoncalvesetal2010}), or there is a relative transverse motion of the jets to the medium
\citep{Bruggenetal2007, Morsonyetal2010, Mendygraletal2012}. These simulations show
that the inflation of bubbles with rapidly precessing jets is similar in many respects to the processes of inflating bubbles with SMW jets.
Instabilities can also prevent the rapid propagation of jets through the ambient gas.  \cite{TchekhovskoyBromberg2015} show how magnetohydrodynamic (MHD) instabilities in AGN jets slow them down and make them deposit their energy in the inner region.

When the jets penetrate to too large distances no fat bubbles are formed, while in intermediate cases elongated bubbles and/or bubbles detached from the center are formed (e.g., \citealt{Basson2003, Ommaetal2004, Heinzetal2006, VernaleoReynolds2006, AlouaniBibi2007, ONeill2010, Mendygraletal2011, Mendygraletal2012}).

Observations in the radio usually present well collimated and narrow jets. The view based on radio observations was that AGN jets are narrow. However this picture has been changing in the last decade, particularly due to the detail and state of the art study conducted by Nahum Arav and his collaborators. The simulations (e.g., \citealt{Sternbergetal2007, Sokeretal2013}) that show that SMW jets can inflate fat bubbles compatible with those observed, and the finding of SMW jets (outflows) in observations seem to fit together into a coherence picture.
SMW jets have been applied also in PNe and are now used in the GEE, and were suggested to occur in young galaxies \citep{Sokeretal2009conf}.
Not all bubbles in cooling flows are inflated by SMW jets, and in many, and possibly all, cases the SMW jets are expected to be accompanied by a faster (relativistic) component that is responsible for the radio emission but carries a small fraction of the energy (e.g. \citealt{Binney2004, Yuanetal2015}).

\cite{Fabian2012} review evidence for outflows from AGN. Most relevant to the present review are the findings of the group of Nahum Arav (e.g., \citealt{Moe2009, Dunn2010, Aravetal2013, Chamberlainetal2015, Liuetal2015, ChamberlainArav2015}), and other similar findings  (e.g., \citealt{Tombesi2012, Harrisonetal2014, Williamsetal2016, Cheungetal2016}) of sub-relativistic wide outflows from different kinds of AGN.
The advantage of their technique is that they can determine the distance from the center where they also calculate the density, and from those and the Doppler blue-shift they calculate the mass outflow rate.

The detection rate of broad absorption line (BAL) outflows from quasars imply that these outflows cover on average a solid angle of $\Omega_j \approx (0.2-0.4)4 \pi$ (\citealt{Chamberlainetal2015} and references therein). \cite{Chamberlainetal2015} list five quasars with energetic outflows, having an outflow velocity range of $4000-10^4 \km \s^{-1}$ and mass outflow rates of about $100-400 M_\odot \yr^{-1}$. In the quasar \astrobj{J0831+03541} they find such a SMW jet at $0.1 \kpc$ from the center. Such SMW jets can inflate fat bubbles \citep{Sternbergetal2007}, and through that transfer energy to the ICM (or ISM) and make the JFM efficient in galactic (e.g., \citealt{Ciottietal2010, Choietal2015}) and cluster (e.g., \citealt{GilkisSoker2012}) scales.
Wide-angle sub-relativistic outflows are revealed by X-ray observation as well (e.g., \citealt{Beharetal2003}).  \cite{Chartasetal2016} for example, find the outflow from the AGN \astrobj{HS~0810+2554} to have a covering factor of $>0.6$, and to carry about ten times more energy than that carried by radiation from the AGN.

Numerical simulations show that jets that are injected with subsonic velocities into the ICM can also inflate fat bubbles (e.g., \citealt{Guo2015, Guo2016}). This goes as well to subsonic jets that are dominated by cosmic rays (e.g., \citealt{GuoMathews2011}). A subsonic jet has a thermal pressure larger than its ram pressure. This implies that the jets expand to the sides at a velocity that is not much smaller than the forward velocity of its head. After a short travel distance the jet will have a large opening angle. Numerically subsonic jets and wide jets are inserted differently, but they can be considered under the same case of wide-angle outflows.

\textbf{$\bigstar$ Summary of subsection \ref{subsec:CFjets}.} Fat-bubbles in cooling flows can be inflated by slow (sub-relativistic), massive, wide (SMW) jets, as well as by precessing and/or jittering jets. SMW outflows (jets) are observed to be blown by AGN.

\subsection{Heating the ICM}
\label{subsec:CFheating}

There is no consensus on the most important processes that transfer energy from the jet-inflated bubbles to the ICM. The processes that were considered favorable until the year 2012 are summarized by \cite{Fabian2012} and \cite{McNamaraNulsen2012}.
\cite{McNamaraNulsen2012} list the following heating mechanisms that were studied (references are given there): shocks, sound waves, and cosmic rays leakage. Another heating process that was discussed is thermal conduction from the outer ICM to the inner cooler regions; this does not involve the AGN.
\cite{Fabian2012} argues that sound waves that are excited by jet-inflated bubbles are more efficient in transferring energy than weak shocks are. Of course, some of the processes can act together, e.g., cosmic rays and thermal conduction (e.g., \citealt{GuoOh2008}).
  I here concentrate on results from the last five years.

\cite{Randalletal2011} and \cite{Randalletal2015} argue, based on their X-ray observations and analysis of the galaxy group \astrobj{NGC~5813}, that shocks excited by periodic jets activity  heat the ICM {{{{ (for early studies of shocks in the \astrobj{Virgo} cluster see \citealt{Formanetal2005} and \citealt{Formanetal2007}). }}}} This heating mechanism is put into question by \cite{SokerHillelS2016}.
Based on deep X-ray observation of the \astrobj{Virgo} and \astrobj{Perseus} cooling flow clusters, \cite{Zhuravlevaetal2014} argue that the main heating process is via dissipation of ICM turbulence. Earlier works on turbulent heating exist (e.g., \citealt{DeYoung2010}), as well as heating by turbulence and turbulent-mixing (e.g. \citealt{BanerjeeSharma2014}). \cite{Falceta2010}, on the other hand, deduce from their numerical simulations that turbulence cannot be the main heating source.
\cite{GilkisSoker2012} and \cite{HillelSoker2014} argue, based on their two-dimensional numerical simulations, that mixing of the hot bubble gas with the ICM is the main heating process.
\cite{BruggenKaiser2002} and \cite{Bruggenetal2009} already discussed mixing of the hot bubble with the ICM as a heating process. They, however, injected artificial bubbles and considered the destruction of the bubble by instabilities. The new simulations show that the mixing takes place by vortices, and the bubble can survive.

\cite{HillelSoker2016} conduct three-dimensional hydrodynamical simulations of jet-inflated bubbles in the ICM and compare the heating by shocks, turbulence, and mixing by vortices.
Three-dimensional simulations are crucial to properly follow the development of vortices in the interface of the bubble and the ICM. These vortices cause mixing and excite turbulence in the ICM.  For the parameters used in those simulations, \cite{HillelSoker2016} find the mixing process to account for approximately $80\%$ of the energy transferred from the jets to
the ICM. About $20 \%$ of the energy that is transferred to the ICM ends in kinetic energy of the ICM. Part of this energy develops as ICM turbulence, and part of this energy is a large scale flow.
They conclude that mixing is the main heating process, with some role played by turbulence,
but only a very small heating by shocks. {{{{ The recent observations of the Perseus cluster with {\it Hitomi} found only weak turbulence in the ICM \citep{Fabianetal2016}, supporting the claims of \cite{HillelSoker2016} based on their 3D numerical simulations. }}}}
\cite{YangReynolds2016b} conduct three-dimensional hydrodynamical simulations for a much longer time. They also conclude that mixing is the main heating process, but find that heating by shocks is the second important process and turbulent heating contributes a small fraction.
Turbulence in the ICM is expected as a by product of the mixing process that is induced by vortices \citep{HillelSoker2016, YangReynolds2016}. This expectation is compatible with findings of turbulence in the ICM of some cooling flows (e.g., \citealt{Zhuravlevaetal2014, Zhuravlevaetal2015, Arevalo2016, AndersonSunyaev2016, Hofmannetal2016}).

\textbf{$\bigstar$ Summary of subsection \ref{subsec:CFheating}.}
Mixing of very-hot bubble gas with the ICM gas seems to be the major channel to heat the ICM. Vortices that are formed in the inflation processes of bubbles cause the efficient mixing.
Vortices also excite turbulence in the ICM. Hence, this scenario is compatible with the finding of turbulence in the ICM of some cooling flows.
The important role of vortices in the interaction of jet-inflated bubbles with the ambient gas is common to other types of systems, e.g., PNe and CCSNe.

\subsection{Closing the feedback cycle}
\label{subsec:CFclosing}

In cooling flows the energy and mass outflow in jets is made possible by mass inflow from the ICM to the SMBH at the center. This inflow closes the AGN-JFM cycle. In some sense the AGN-JFM feedback is not only a feedback cycle of energy, but also of mass.

There are several lines of evidence for the non-continues operation of the JFM in cooling flows. Namely, there are periods when no jets are launched (e.g., \citealt{Werneretal2016}). A strong evidence for cyclical JFM comes from clusters where two or more bipolar pairs of bubbles are seen, such as \astrobj{Perseus} (e.g. \citealt{Fabianetal2000}) and \astrobj{Hydra~A} (e.g., \citealt{Wiseetal2007}).
In some other types of systems reviewed here the launching processes might be continues, but a cyclical variation might take place because of an orbital motion, such that the location of the jet-launching changes; such is the case in the CEE, GEE, and ILOTs.

Two modes of accretion onto the SMBH are considered in the literature. One is accretion of the hot ICM gas, i.e., a Bondi-type accretion. The other one is accretion of cold clumps of gas, with temperatures of $T\la 10^4 \K$, that cooled from thermal instabilities in the ICM.
In principle both modes can work in the feedback cycle (e.g., \citealt{Gasparietal2011}).
While Bondi accretion appears to be energetically plausible in cooling flows in elliptical galaxies (e.g., \citealt{Allenetal2006, NarayanFabian2011, GuoMathews2014}), it suffers from severe problems in cooling flows in clusters of galaxies (e.g., \citealt{Raffertyetal2006, Sokeretal2009conf, McNamaraetal2011}).

The process of feeding the AGN with cold clumps as part of the AGN-JFM is termed the
\emph{cold feedback mechanism}, and was suggested by \cite{PizzolatoSoker2005a} and is furthered developed by \cite{PizzolatoSoker2010}.
{{{{ The cold clumps can survive down to very close to the AGN (e.g., \citealt{Kuncicetal1996}. }}}}
As the cold clumps are formed through thermal instabilities, many of their properties, such as velocity distribution, have chaotic distribution. \cite{Gasparietal2013} term this processes chaotic cold accretion, and \cite{Gaspari2012b} discuss a `rain' of cold clumps to the center (also \citealt{Gasparietal2015} and \citealt{Gaspari2015arXiv}). \cite{VoitDonahue2015} attach the name `precipitation' to such a `rain' of cold clouds.
\cite{PizzolatoSoker2005a} and \cite{Soker2006aNewA} further suggest that jets and bubbles form the perturbations that are required to form the cold clumps for future feeding cycles. A process along this suggestion is further developed by \cite{Revaz2008}, \cite{Li2014b}, and \cite{McNamaraetal2016s}, who consider the perturbations to develop from low entropy gas that is uplifted by the rising bubbles. \cite{McNamaraetal2016s} term this process the stimulated AGN feedback.

Overall, many observational and theoretical studies in the last decade have put the cold feedback mechanism on a very solid ground (e.g., \citealt{Revaz2008, Pope2009, Wilmanetal2009, Edgeetal2010, Wilman2011, Nesvadba2011, Cavagnolo2011, Gaspari2012a, Gaspari2012b, McCourt2012, Sharmaetal2012MN427, Sharma2012, Farage2012, Waghetal2013, Li2014a, Li2014b, McNamaraetal2014, VoitDonahue2015, Voitetal2015b, Lietal2015, Prasadetal2015, SinghSharma2015, Tremblayetal2015, ValentiniBrighenti2015, ChoudhurySharma2016, Hameretal2016, Loubseretal2016, Meeceetal2016, Russelletal2016, McNamaraetal2016s, YangReynolds2016b, Baraietal2016, Tremblayetal2016}).
As an example, \cite{Gaspari2015MN451} shows that the AGN-JFM self-regulated by accretion of cold clumps can explain the strong deficit of soft X-ray emission in clusters and groups of galaxies.
As well, there are hints of a cold feedback mechanism operating in young galaxies (e.g.,
\citealt{Colletetal2016}).

\cite{Gaspari2012b} and \cite{Prasadetal2015} conduct 3-dimensional numerical simulations of the AGN-JFM in cooling flows, covering the entire feedback cycle. In both studies cold clumps feed the AGN, i.e., the cold feedback mechanism, and in both studies sub-relativistic jets carry the energy and mass from the AGN to the ICM. To prevent jet-penetration, \cite{Gaspari2012b} use jittering (random wobbling) jets, and \cite{Prasadetal2015} inject wide-angle jets (a half opening angle of $\alpha_j=30^\circ$).
\cite{Lietal2015} also simulate full AGN-JFM cycles, using sub-relativistic precessing jets,
and the cold feedback mechanism to close the cycle.
\cite{YangReynolds2016} conduct 3-dimensional magnetohydrodynamic simulations. In some of the simulations they inject slow-precessing jets, and consider the cold feedback mechanism.
These high quality simulations nicely demonstrate the full cycle of the AGN-JFM in cooling flows.

The cold feedback mechanism might lead to a chaotic accretion. Namely, the mass and angular momentum accretion rates, as well as the angular momentum axis, change with time, in some cases in an erratic manner. This process can cause a change in the jets' axis from one episode to the next.
\cite{Schlegeletal2016} observe what might be episodic outbursts with changing directions, and argue for an evidence for an AGN feedback in the galaxy \astrobj{NGC 5195}.
\cite{KingPringle2006} discuss chaotic accretion of angular momentum onto the SMBH in young galaxies. In section \ref{sec:SNe} I discuss erratic accretion onto the central object that forms when massive stars explode in CCSNe, a processes that might lead to jittering jets.
\cite{Babuletal2013} suggest that the SMBH at the center of cooling flows might experience short stochastic episodes of enhanced accretion via thin disks, a process that when combined with the spin of the black hole leads to reorientation of the axis of the jets, i.e., jittering jets.

\textbf{$\bigstar$ Summary of subsection \ref{subsec:CFclosing}.}
It seems that cold clumps of gas are the main source of mass that feeds the SMBH in the AGN of cooling flows. This is termed the cold feedback mechanism. Many observations from the last decade indicate the presence of large enough quantities of cold clouds near the center of cooling flows to feed the AGN. Theoretical studies, including numerical simulations of full feedback cycles, strongly support the cold feedback mechanism.

\section{Galaxy Formation}
\label{sec:galaxy}

There is a large body of literature on feedback processes during the epoch of galaxy formation, including feedback from stars, mainly through supernovae, and from AGN, by radiation and/or jets/outflows (e.g., \citealt{Cattaneoetal2009, AlexanderHickox2012, Fabian2012, KormendyHo2013, HeckmanBest2014, KingPounds2015, Tadhunter2016} for reviews, and, e.g., \citealt{Wagneretal2016, Bongiornoetal2016, Weinbergeretal2016} for recent papers on the AGN-JFM). In this section I review only the AGN-JFM, and concentrate only on processes that are related to other objects that are reviewed here, in particular to cooling flows in clusters of galaxies.

Some reviews list the evidences and hints to the operation of an AGN-JFM during galaxy formation and growth (e.g. \citealt{KormendyHo2013}), and some discuss the similarities of this feedback to the feedback process in cooling flows. One very strong motivation to discuss AGN feedback during galaxy formation comes from the correlations found between the mass of the central SMBH and properties of the elliptical galaxy, or the bulge in spiral galaxies (e.g., \citealt{AlexanderHickox2012, KormendyHo2013, KingPounds2015}). I take the stand that mergers of small galaxies to build larger galaxies cannot account for these correlations \citep{Ginatetal2016}. As well, it seems that AGN radiation is not capable to account for this correlation (e.g., \citealt{SilkNusser2010}). If these correlations are determined by a feedback mechanism, we are left therefore with the AGN-JFM to account for these correlations. If holds, this allows the usage of knowledge acquired from the AGN-JFM operating in cooling flows to infer about the AGN-JFM that has been operating during galaxy formation and growth. Since observations of the feedback cycle that takes place during galaxy formation are less informative than those of cooling flows in clusters of galaxies, the comparison of the two environments can teach us on the formation processes of galaxies.
Another view is that a common gas supply is the primary driver of the observed correlations, rather than a feedback mechanism (e.g., \citealt{Anglesetal2016}).

Possible similarities between the two different AGN-JFM environments are the properties of the outflows from the AGN, in particular sub-relativistic outflows (e.g., \citealt{KormendyHo2013, KingPounds2015}). In addition to the evidence in the visible and UV bands for slow-massive-wide outflows from AGN discussed in section \ref{subsec:CFjets} (e.g., \citealt{ChamberlainArav2015, Zakamskaetal2016}), there are indications in the X-ray band for sub-relativistic and relatively massive outflows from AGN (e.g., \citealt{Beharetal2003, Poundsetal2003, Tombesietal2010, Chartasetal2016}). Such outflows can efficiently interact and influence the ambient gas (section \ref{subsec:CFheating}), in particular via the inflation of hot low density bubbles.
The discovery of bipolar lobes even in spiral galaxies (e.g., \citep{Singhetal2015, Querejetaetal2016}, including in the Milky Way \citep{Suetal2010} that are likely to have been inflated by jets \citep{GuoMathews2012, Guoetal2012, Mouetal2015}, and possibly in Andromeda \citep{Pshirkovetal2016}, suggests that the feedback process in galaxies, even on scales of the bulge of spiral galaxies, might take place through the inflation of bubbles.

There are some differences between the AGN-JFM in cooling flows in clusters of galaxies and during galaxy formation. The first one is that in cooling flows in clusters of galaxies the main role of the jets is to heat the ICM, while during galaxy formation the jets are expected to expel gas from the young galaxy (e.g., \citealt{Boweretal2008}), {{{{ e.g., as observed in some young radio galaxies (e.g., \citealt{Holtetal2008}). \cite{Duboisetal2016}, as another example, claim that AGN feedback is essential in order to produce massive galaxies that resemble elliptical galaxies.  }}}}

In cooling flows gas cooling from the hot ICM is the main source that feeds the AGN. Such a process might take place during galaxy formation as well. It implies that a cooling flow phase might occur during galaxy formation \citep{Soker2010MN407}. \cite{Guo2014} further mentions the possibility that some properties of galaxies are determined through a cooling flow phase, and \cite{Voitetal2015a}  show that a feedback cycle operating similarly to that in cooling flows can explain some observed scaling relationships among global properties of a galaxy.
\cite{Colletetal2015} conclude from their observation of cold gas in two galaxies that a processes similar to cooling flows in clusters of galaxies is operating in these two galaxies.
\cite{KormendyHo2013} mention that AGN feedback prevents galaxy formation from ``going to completion'' by keeping baryons locked up in hot gas. They basically refer to a situation like in a cooling flow. As jets interact much more efficiently with the diffuse hot gas than with cold clumps, it might be that the main feedback process at galaxy formation takes place during a cooling flow phase, and operates as a cold feedback mechanism that leads also to the formation of stars. This deserves further study.
Although a cooling flow phase might occur at galaxy formation, it might be that this is not the main feeding source of the center of the galaxy and its AGN. Most of the mass might flow inward from the inter-galactic medium to the galaxy in cold streams (e.g., \citealt{DekelBirnboim2006}).

There are indications of time variability in AGN activity during star formation (e.g., \citealt{Hickoxetal2014}). As well, accretion of mass on to the SMBH can be chaotic and lead to a feedback mechanism with jets having chaotic variations of their direction \citep{KingNixon2015}. This is expected also to occur in the cold feedback mechanism discussed in section \ref{subsec:CFclosing} \citep{Gasparietal2013}, and in the jittering-jets model of supernovae to be discussed in section \ref{sec:SNe}. Jets with chaotically varying directions (jittering jets) interact more efficiently with the ambient gas.

There is no consensus on whether the interaction of the jets with the ISM during galaxy formation is dominated by a momentum conserving or an energy conserving case.
The large ratio of SMBH potential to that of the galaxy, $\Phi_a/\Phi_{\rm res} \approx 10^6$
(Table 1), implies that the transfer of energy from the accreted gas to the ISM may be inefficient and still substantially influence the thermal and kinematical states of the ISM. We will encounter such a situation in the case of young stars (Section \ref{sec:YSOs}).
\cite{Ostrikeretal2010} find that the momentum and mass carried by jets are important for the feedback cycle. This requires SMW jets, as found by, e.g., \cite{Moe2009}.
\cite{KingPounds2015} suggest that the feedback changes from being momentum driven close to the center, to being energy driven on a larger scale.
\cite{SilkNusser2010} argue that the AGN cannot supply the required momentum in radiation to account for the correlation of the SMBH mass with its host properties.
\cite{Costaetal2014} conduct hydrodynamical numerical simulations and conclude that the interaction that leads to massive outflows from galaxies is energy conserving.
{{{{ \cite{Duganetal2016} simulated both narrow jets and SMW jets launched at the center of a gas-rich disk galaxy. They find that narrow jets feedback is energy-driven, while the SMW jet feedback is momentum-driven.
In any case, turbulence in the ISM helps in achieving the non-penetrating condition, hence in retaining more of the jets energy, leading more easily to bubble inflation (e.g., \citealt{Mukherjeeetal2016}). }}}}

{{{{ I end by raising a speculative possibility about the first time an AGN-JFM is turned on. I suggest a scenario where the JFM operates continuously from a young (stellar) object (YO) through a SN, and then in a newly born AGN. The formation of SMBH at high red-shifts might be accounted for by the core collapse of a massive star (e.g., see a recent  review by \citealt{LatifFerrara2016}). The proposed YO$\rightarrow$SN$\rightarrow$AGN (YOSA) JFM scenario is possible only if the first SMBHs were formed from such supermassive stars. The evolution is very rapid, and the collapse can take place while the accretion process from the ISM to the center did not terminate yet. The mass accretion rate is $\ga 0.1 M_\odot \yr^{-1}$. With a YO mass of $M_{\rm YO} \approx 10^6 M_\odot$ and a radius of $R_{\rm YO} \approx 10^3 R_\odot$ (e.g., \citealt{Begelman2010}), the escape velocity from this star is $\approx 2 \times 10^4 \km \s^{-1}$. The jets (or disk wind) from the accretion disk around such an object can release an energy of $\approx 10^{56}-10^{57} \erg$, equivalent to $\approx 10^5-10^6$ CCSNe. The jets affect the cloud from which the supermassive star was formed in a feedback mechanism. After the center of the supermassive star collapses, an accretion disk is formed around the newly born BH, and jets explode part of the star, mainly along the polar directions. Neutrino cooling is not important and cannot account for such an explosion. The explosion is determined by the JFM, but material stays bound near the equatorial plane and continues to feed the newly born SMBH. In this speculative YOSA-JFM scenario, the AGN-JFM is born directly from a CCSN-JFM that itself is a continuation of a YSO-JFM. }}}}

\textbf{$\bigstar$ Summary of section \ref{sec:galaxy}.}
The main conclusion one can draw from this section is that a cooling flow phase, similar to cooling flows in clusters of galaxies in the nearby Universe, might have taken place during the formation period of galaxies.

The AGN-JFM operating in both environments has similarities. But it has some differences.
For most, during galaxy formation feedback from stars (supernovae) and radiation from the AGN play a non-negligible role, unlike in cooling flows in clusters of galaxies where jets play the dominant role.
Hence, although the AGN-JFM seems to operate in galaxy formation, it is not as crucial as in cooling flows. The AGN-JFM might be the dominant factor in determining the correlations between the mass of the SMBH and some galactic properties.

Another relevant property of the AGN-JFM is that stochastic variations in the angular momentum of the accreted gas might lead to the launching of jets with stochastically varying directions, i.e., jittering jets. Jittering jets will be discussed in the next section in relation to supernovae. It seems that dense clumps of gas feed the compact objects in some of the systems studied in this review, a process than might lead to stochastic properties of the jets.

There is no consensus yet of whether the interaction of the jets with the ISM is through a momentum conserving process, where radiation carries most of the energy of the post-shock jets' gas, or whether the interaction is energy conserving. It is my view, based on the similarity with cluster cooling flows, that the interaction is energy conserving through the inflation of bubbles.

\section{Core collapse supernovae (CCSNe)}
\label{sec:SNe}
\subsection{Motivation}
\label{subsec:SNe:motivation}
Whether a JFM operates or not in CCSNe is in a fierce dispute. However, there are three lines of arguments that have motivated the development of the JFM for CCSNe.

(1) \emph{Explosion energy.} Observations show that most CCSNe explode with a typical energy, kinetic energy plus radiation energy, that is about equal to, or just several times, the binding energy of the ejected mass, $E_{\rm explosion} \simeq {\rm few} \times E_{\rm bind} \approx 10^{51} \erg$. Most of the binding energy is of the material ejected from the stellar core. This hints at the operation of a negative feedback mechanism. Namely, as long as the energy in the explosion is not sufficient to eject the core, the engine continues to act. The JFM has this property.

(2) \emph{Jets in GRBs that come with SN Ic.} Long gamma ray bursts (GRBs) are associated with Type Ic supernovae (e.g., \citealt{WoosleBloom2006, Canoetal2016}). It is now commonly accepted that long GRBs are powered by jets launched from a NS or a BH (e.g., \citealt{Woosley1993, ShavivDar1995, SariPiran1997}).
This demonstrates that jets can be produced from the collapse of the inner core of massive stars. The question is whether jets are formed only after the explosion, or the jets cause the explosion.
Here the view that jets cause the explosion is adopted.
To that I add the accumulating evidences for a real jet in SN remnant \astrobj{Cassiopeia~A}
\citep{FesenMilisavljevic2016}, and observations (e.g. \citealt{Milisavljevic2013, Lopez2013})  suggesting that jets might play a role in at least some CCSNe.

(3) \emph{Problems with the delayed neutrino mechanism.}
The most studied CCSN explosion mechanism in the last three decades, starting with \cite{Wilson1985} and \cite{BetheWilson1985}, is the delayed neutrino mechanism (see, e.g.,
\citealt{Janka2012} and \citealt{Jankaetal2016} for reviews).
There are three obstacles that the delayed neutrino mechanism should overcome \citep{Papishetal2015}.
In the delayed-neutrino mechanism the explosion starts with the revival of the stalled shock of the inflowing core gas. Namely, the energy deposited by neutrinos in the region inward to the stalled shock must revived the shock. This is not always achieved in numerical simulations, even the most sophisticated ones.
Even if the stalled shock is revived, in most simulations the desired energy of $\approx 10^{51} \erg$ is not achieved.
It is my view that the delayed-neutrino mechanism has failed to consistently and persistently overcome these two obstacles. This is evident from the varying, and sometimes conflicting, results of increasingly sophisticated 2-dimensional and 3-dimensional simulations \citep{Papishetal2015}.
\cite{Papishetal2015} argue that even if the stalled shock is revived, there is a generic problem that limits the explosion energy in the delayed-neutrino mechanism to $E_{\rm explosion} \la 0.2-0.5 \times 10^{51} \erg$.

The third obstacle is of a different kind. When in simulations based on neutrino-driven explosion the explosion energy is scaled to observed CCSNe, such as \astrobj{SN~1987A}, the maximum energy that the delayed-neutrino mechanism can supply is about $2 \times 10^{51} \erg$
(e.g., \citealt{Sukhboldetal2016}), and at most $3 \times 10^{51} \erg$ \citep{SukhboldWoosley2016}. Even when convection-enhanced neutrino-driven explosion is considered (convective-engine) the explosion energy cannot get higher than this limit \citep{Fryer2006, Fryeretal2012}.
The delayed neutrino mechanism cannot account for super energetic CCSNe (SESNe), such as \astrobj{ASASSN-15lh} with an explosion energy of $E_{\rm explosion} > 10^{52} \erg$ \citealt{Dongetal2016}). This demonstrates the limitation of neutrino-based mechanisms. \cite{Kushnir2015b} lists some other difficulties of the delayed neutrino mechanism.

\subsection{The jittering-jets model}
\label{subsec:SNe:jittering}

For the lack of a persisting success, and possibly even a failure, of the delayed-neutrino mechanism, and the supporting observations for the presence of jets and hints on a possible operation of a negative feedback mechanism, it was suggested that all CCSNe with explosion energies above about $0.2 \times 10^{51} \erg$ are exploded via a JFM.

Jet-driven explosions of CCSNe have been simulated for a long time, but mainly in cases where the pre-collapsing core possesses both a rapid rotation and strong magnetic fields (e.g.
\citealt{LeBlanc1970, Meier1976, Bisnovatyi1976, Khokhlov1999, MacFadyen2001, Hoflich2001, Woosley2005, Burrows2007, Couch2009, Couch2011, TakiwakiKotake2011, Lazzati2012, BrombergTchekhovskoy2016}).
However, these jets can be formed only in rare cases of rapid core rotation, they do not operate via a feedback mechanism, such that they too often give extreme cases as gamma ray bursts, or
they fail to explode the star, e.g., \cite{Mostaetal2014}.

The challenge then is to supply the required angular momentum to form an accretion disk around the newly born NS of BH. To overcome this challenge the jittering-jets mechanism was constructed \citep{Soker2010CCSN, PapishSoker2011, PapishSoker2012, PapishSoker2014a, PapishSoker2014b}, supplied with the idea that accretion belts can launch jets \citep{SchreierSoker2016}.
The angular momentum source is the convective regions in the pre-collapse core \citep{GilkisSoker2014, GilkisSoker2015, GilkisSoker2016}, and, more likely, instabilities in the shocked region of the collapsing core, such as the standing accretion shock instability (SASI; \citealt{Papishetal2016}).
Relevant to the stochastic angular momentum supply are the spiral modes of the SASI (for more on the spiral modes of SASI see, e.g., \citealt{Hanke2013, Fernandez2015, Kazeronietal2015}). In that respect it should be noted that \cite{BlondinMezzacappa2007}, \cite{Fernandez2010}, and \cite{Rantsiouetal2011} suggest that the the SASI can supply the angular momentum of pulsars.
New simulations suggest that the SASI is the main source of the angular momentum, and not pre-collapse convection, although the convection can supply the seed perturbations.

\begin{figure}[ht!]
   \centering
    \includegraphics*[scale=0.42]{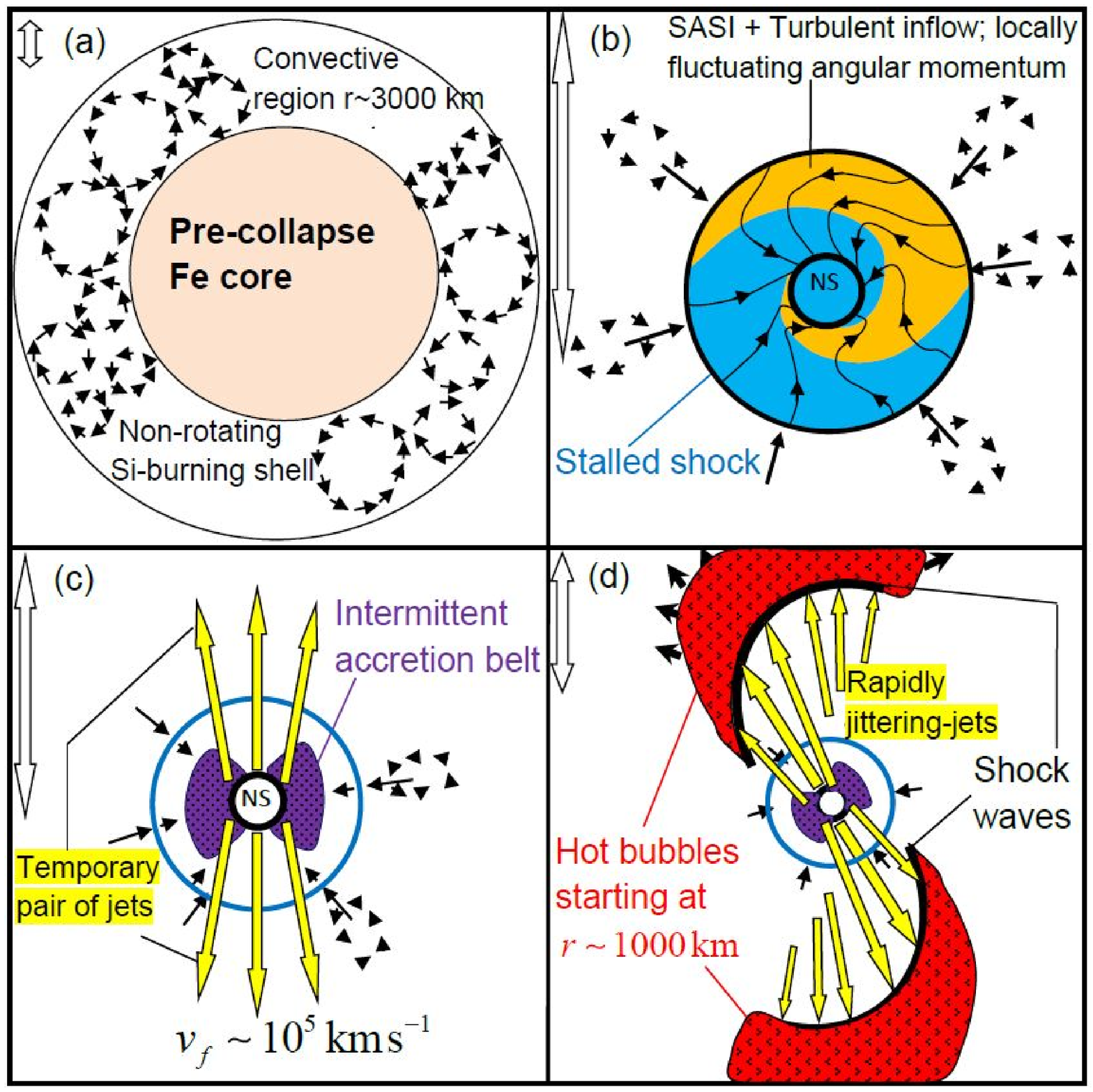} \\
\caption{A schematic presentation of the jittering-jets mechanism for the case of a non-rotating (or slowly rotating) core. The evolution is several seconds (taken from \citealt{Papishetal2016}). The two-sided arrow on the upper left of each panel corresponds to a length of approximately $500 \km$.
(a) Emphasizing the convective vortices in the silicon burning shell in the pre-collapse core. These serve as the source of stochastic accreted angular momentum. (b) After the pre-collapse NS formation the in-falling gas flows through the stalled shock. Further stochastic angular momentum of the accreted gas is introduced by the spiral modes of the standing accretion shock instability (SASI).
(c) For short periods of times, tens of milliseconds, the stochastic angular momentum of the accreted gas leads to the formation of an intermittent accretion belt around the newly born NS. This belt might spread to form a disk.  The belt or disk are assumed to launch two opposite jets with initial velocities of $v_f \approx 10^5 \km \s^{-1}$. (d) The jittering-jets inflate hot bubbles. These bubbles expand and explode the star \citep{PapishSoker2014a, PapishSoker2014b}.}
      \label{fig:schem1}
\end{figure}

These angular momentum sources are stochastic in nature. Hence, the accretion disk or belt axis changes in a stochastic manner, and so is the direction of the jets. This launching process is termed the jittering-jets mechanism, and it is presented schematically in Figure \ref{fig:schem1}.
The intermittent accretion disk, or accretion belt, launch jets in periods, each lasting for about $0.01-0.1 \sec$. The changing direction prevents the jets from drilling out of the core. The non-penetrating condition for the jets to deposit their energy in the core, a radius of thousands of km, was driven by \cite{PapishSoker2011}.
In the jittering-jet mechanism there is no need to revive the accretion shock, and it is a mechanism based on a JFM. Namely, as long as the core was not exploded, the accretion continues. In cases when the JFM is efficient, then after an energy equals several times the core binding energy is deposited to the core by the jets, the star explodes.  This energy amounts to $\approx 10^{51} \erg$.

{{{{ \cite{Maundetal2007} find departure from axial symmetry from the polarization in the
Type Ib/c SN 2005bf. They  suggest that a jet whose axis is tilted with respect to the axis of the photosphere, and is rich in Nickel-56 has penetrated the C–O core, but not the He mantle. Such  jets are naturally accounted for in the jittering jet mechanism. }}}}

{{{{ Some SN remnants (SNR) have morphologies that suggest they have been shaped by jets. Such is the structure of two opposite `ears', as in the SNR Puppis A. \cite{Castellettietal2006} suggest that Puppis A was shaped by jets. Here I add that such jets might be the two opposite jets launched at the last jets-launching episode in the jittering jets model, after the core has been exploded. The jets might expand more freely after the core has been removed, and shape the SNR.  }}}}

\subsection{Accretion belts}
\label{subsec:SNe:belt}

The question whether the pre-collapse convection and the post-collapse instabilities can supply sufficient angular momentum is an open one. In many cases the accreted gas might have a specific angular momentum lower than what is required to form an accretion disk around the accreting object \citep{Papishetal2016, GilkisSoker2016}.
This sub-Keplerian inflow, therefore, cannot form a rotationally supported accretion disk, at least not initially. The gas streams onto the accreting body from a wide solid angle around the equatorial plane. Due to the non-zero specific angular momentum there are two avoidance regions, one center at each pole, where the accretion rate is very low. \cite{SchreierSoker2016} suggest that a turbulent region develops where the accretion belt interacts with the accreting object via a shear layer. They further speculate that a magnetic dynamo that is developed in the shear layer amplifies the magnetic fields, and the magnetic fields in turn form polar outflows from the avoidance regions.

The density in the accretion belt is very high in the relevant cases reviewed here, and in general there is no time for radiation to escape, even not for neutrinos in the case of CCSNe, during several dynamical times. Further accretion of mass from further out is made possible as the jets themselves remove energy and angular momentum from the surface of the accreting body. This raises the possibility that accretion belts might actually be more efficient than accretion disks in producing jets. The testing of this speculation requires high-resolution 3-dimensional MHD simulations. The possibility that accretion belts efficiently launch jets has implications to the CEE with a main sequence companion (section \ref{sec:CEE}).

The basic ingredients of the belt-launched jets scenario are presented in Table 2 alongside with those of jets launched by accretion disks. The belt-launched jets scenario does not require large scale magnetic fields and does not require a thin Keplerian disk. The most important ingredient is the operation of a dynamo, the belt-dynamo \citep{SchreierSoker2016}.
\begin{table}[ht]
 \centering
 \begin{tabular}{l c c} 
 \hline
 Physical parameter      &      Accretion disk          &  Accretion Belt  \\
 [0.5ex]
 \hline \hline
 Accretion rate ($\dot M_{\rm acc}$)&up to $\approx \dot M_{\rm Edd}$& $\ga \dot M_{\rm Edd}$  \\
 \hline
Accreted specific angular& $ \gg j_{\rm Kep}$           &   $ < j_{\rm Kep}$  \\
momentum ($j_{\rm acc}$) &                              &      \\
 \hline
Angular velocity         & $\approx \Omega_{\rm Kep}$   & $< \Omega_{\rm Kep}$   \\
at launching ($\Omega_ {\rm L}$)  &                     &      \\
 \hline
 Launching area ($D_{\rm L}$)  &  $\gg R_a$               &    $\simeq R_A$ \\
 \hline
 Magnetic fields        & weak to strong   & strong; $ \vec B$ amplified       \\
                        &                  & locally via dynamo       \\
 \hline
 Radiative        & Significance     & Might be small        \\
 losses           &                  & at high $\dot M_{\rm acc}$       \\
 \hline
 Jet's energy          & Gravitational of     &  Gravitational of \\
  source               & accreted gas         &  accreted gas     \\
  \hline
  \end{tabular}
\label{table:belt}t
\newline
\caption{Comparison of the belt-launched jets scenario with that of jets launched by a thin accretion disk.
 The different symbols have the following meaning: $R_a$ is the
 radius of the accreting body; $\dot M_{\rm Edd}$ is the Eddington mass accretion rate limit; $j_{\rm Kep}$ and $\Omega_{\rm Kep}$ are the Keplerian specific angular momentum and angular velocity very close to the accreting body surface, respectively (based on \citealt{SchreierSoker2016}).
 }
\end{table}

\subsection{Inefficient JFM}
\label{subsec:SNe:ineffcient}

\begin{figure}[ht!]
   \centering
   \includegraphics*[scale=0.42]{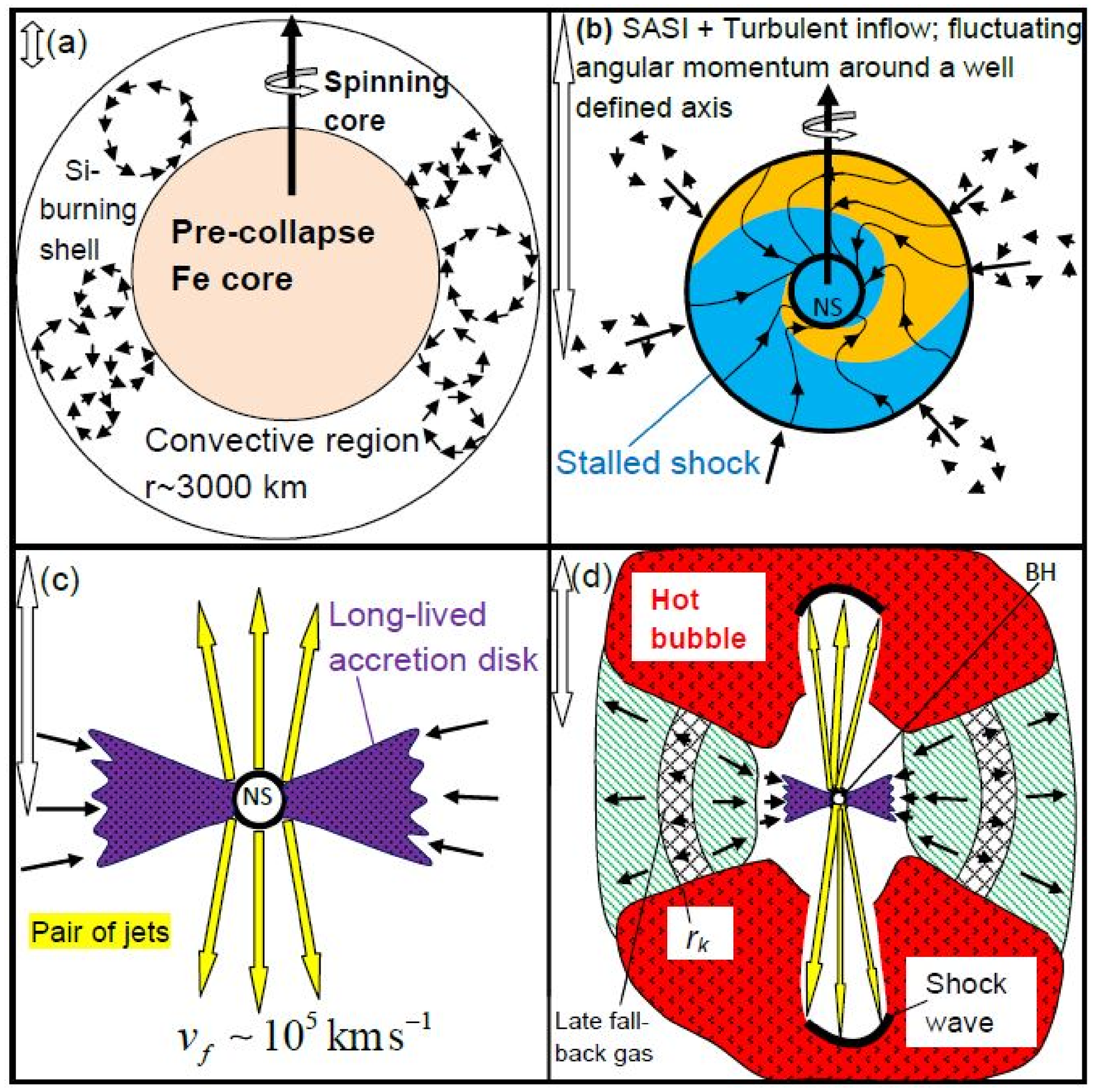} \\
\caption{A schematic presentation of an inefficient JFM that takes place when the pre-collapse core is rapidly rotating (taken from \citealt{Gilkisetal2016}).
 (a) A rapidly spinning pre-collapse core serves as the main source of angular momentum of the accreted mass at later times.
 (b)+(c) Because of the large specific angular momentum of the accreted gas a long-lived accretion disk is formed around the newly born NS or BH.  The jets preserve their axis, with small jittering due to convection and angular momentum. In the first three panels the two-sided arrow on the upper left represent a length of about $500 \km$.
 (d) The jet-inflated bubbles (in red) expel mass from two large regions to the sides of the equatorial plane. This removal of mass reduces the gravitational force on the gas in the equatorial plane vicinity (marked by hatched green region), and its outer parts start to move outward due to the pressure gradient. The mass in the inner parts continues to flow inward as a rarefaction wave propagates out. The mass in the two cross-hatched arcs in the figure at radius marked $r_k$ (a ring in 3D), is accelerated outward at early times, but not to the escape speed. In a time much longer than the dynamical time, this mass might fall back and power jets minutes to days after explosion, and when a BH has been already formed. In this panel the two-sided arrow on the upper left corresponds $\approx 0.1 R_\odot$. }
      \label{fig:schem2}
\end{figure}

If the feedback is less efficient, more accretion is required to accumulate the required energy to eject the rest of the core. If the efficiency is very low, the accreted mass onto the NS brings it to collapse to a black hole and launch relativistic jets.  Namely, in general, \emph{the less efficient the feedback mechanism is, the more violent the explosion is} \citep{GilkisSoker2014}. When the pre-collapse core is rapidly rotating, the jets will be well collimated and not efficient in removing mass from the equatorial plane vicinity. This case was studied by \cite{Gilkisetal2016} and is presented schematically in Figure \ref{fig:schem2}.

\cite{Gilkisetal2016} suggest a scenario where an inefficient JFM operating in the collapse of a rapidly rotating core accounts for super-energetic supernovae (SESNe), such as \astrobj{ASASSN-15lh} \citep{Dongetal2016}. The jets that are launched by the central NS (or BH) maintain their direction and remove mass along the polar directions, but not from near the equatorial plane (see Figure \ref{fig:schem2}). The mass removal reduces the gravity near the equatorial plane, and mass is accelerated outward due to the pressure gradient. There is also mass removal by neutrinos (e.g., \citealt{Nadezhin1980, GoldmanNussinov1993, LovegroveWoosley2013}), but the mass removal by jets in this scenario is more significant. Part of the mass might fall back over a timescale of minutes to days to prolong the jets-launching episode, and possibly power a SESN. \cite{Gilkisetal2016} speculate that the mass that escapes forms a slow equatorial outflow, with typical velocity and mass of $v_\mathrm{eq} \approx 1000 \kms$ and $M_\mathrm{eq} \ga 1 M_\odot$, respectively.

According to the JFM scenario for CCSNe, the principal parameter that determines the outcome of the CCSN explosion, such as whether it be a normal CCSN or a SESNe, is the efficiency of the JFM. This efficiency in turn depends mainly on the pre-collapse angular momentum profile in the core, and to lesser degree on the pre-collapse core mass, envelope mass, and core convection.

One alternative model for powering SESNe is based on strongly-magnetized, rapidly-rotating NS, termed magnetars. Magnetars have been suggested also to power long-duration GRBs (e.g., \citealt{Metzgeretal2015} for a recent paper, {{{{ and \citealt{UzdenskyMacFadyen2007} for a model with a late fallback). }}}}
In the magnetar model the CCSN explosion that occurs when the NS is formed is power by the delayed neutrino mechanism.
However, \cite{Soker2016c} claims that under reasonable assumptions jets are expected to be launched when a magnetar is formed. This claim is based on studies that show that when magnetized gas with high specific angular momentum is accreted onto a NS, the magneto-rotational instability (MRI) and dynamo lead to the launching of a bipolar outflow \citep{Akiyamaetal2003, Mostaetal2015}, and on the belt-launched jet model \citep{SchreierSoker2016}.
\cite{Soker2016c} further estimates that the kinetic energy that is carried by the jets does not fall much below, and might even surpasses, the energy that is stored in the newly born spinning NS. This brings us back to the JFM.
As well, some fraction, probably a small fraction, of the rotational energy of the NS might be channelled to gravity waves \citep{MoriyaTauris2016}.
{{{{ More generally, the rapid rotation that is required to form the magnetar can lead to energetic explosion before the magnetar powers the explosion, not only via the MRI mechanism. \cite{Blackmanetal2006} show that both magnetic fields and turbulence dissipation can extract energy from differential rotation and rotation, and power an explosion. To conclude, the formation process of a magnetar cannot be ignored when energy supply by a magnetar is discussed.  }}}}

I also note that the rapid core rotation required in the collapse-induced thermonuclear explosion (CITE) mechanism for CCSNe model \citep{Burbidgeetal1957, Kushnir2015a} is very likely to lead to the launching of very energetic jets that dwarf the energy released by the thermonuclear burning \citep{Gilkisetal2016}.

\textbf{$\bigstar$ Summary of section \ref{sec:SNe}.} The main points raised in this section and their relation to other systems reviewed here can be summarised as follows.
\newline
(1) The notion that CCSNe are exploded by neutrinos is deeply rooted in the community. However, a critical examination of results in recent years shows that the delayed neutrino mechanism did not yield explosions with the desired energy. Alternative explosion mechanisms should be examined.
I call for a paradigm shift, from neutrino to jets driven explosions.
   \newline
(2) As in other systems studied here, the jets work in a negative feedback cycle. This is the JFM.
    \newline
(3) In the common CCSNe, those with explosion energies of $E_{\rm explosion} \simeq  10^{51} \erg$, the jets have a stochastically varying direction. This is termed the jittering-jets model. The JFM works efficiently to explode the core and envelope of the star and a NS is formed. The jittering ensures the non-penetrating condition for the jets, as discussed for other objects, e.g., in section \ref{sec:CFs}
 \newline
(4) When the pre-collapse core is rapidly rotating, the axis of the jets does not change direction. The jets do not remove mass from the equatorial region, and the JFM is less efficient. More mass is accreted onto the central object, a NS and then a BH. As a result more mass is launched in the jets and the event might be more energetic. Super-energetic CCSNe might be formed from the operation of an inefficient JFM.
   \newline
(5) The main explosion parameter that distinguishes between the different cases and energies of CCSN explosions is the efficiency of the JFM.
   \newline
(6) As in the case of the CEE discussed next, the operation of the jittering-jets model requires that accretion belts launch jets. This speculative idea has far reaching consequences, and should be studied in details.

I end this section with a look into the future.
If 3-dimensional simulations of CCSNe in the frame of the delayed neutrino mechanism will not achieve explosions with the desired energy of about $10^{51} \erg$ by the year 2020, the delayed neutrino mechanism should be abandon as a general explosion mechanism. The JFM, including the jittering-jets model, that is still in a speculative phase, will become the most promising alternative.
Already now the JFM does better than the delayed neutrino mechanism in explaining CCSNe with explosion energies of $>2 \times 10^{51} \erg$.

\section{Common Envelope Evolution (CEE)}
\label{sec:CEE}

The motivation to consider jets launched by the compact companion as it spirals inside the envelope of the giant star is twofold.
Firstly, there are still problems to eject the common envelope in numerical simulations that do not consider jets \citep{Soker2013}. Secondly, \cite{BlackmanLucchini2014} deduce from the momenta of bipolar pre-PNe that strongly interacting binary systems, probably in a common envelope, can launch energetic jets.
But it should be kept in mind that the CEE does not require the operation of the JFM. The jets are not crucial, and might operate only in a fraction of all CEE cases.

\cite{BodenheimerTaam1984} conducted 2-dimensional hydrodynamical simulation of the CEE, followed by 3-dimensional simulations by \cite{LivioSoker1988} who used a particle in cell (PIC) hydrodynamical code.
More simulations followed over the years using grid-based codes or smoothed-particle hydrodynamics (SPH) codes  (e.g., \citealt{RasioLivio1996, SandquistTaam1998, Lombardi2006, RickerTaam2008, TaamRicker2010, DeMarcoetal2011, Passyetal2011, Passyetal2012, RickerTaam2012, Nandezetal2014, Ohlmannetal2016, Iaconietal2016, Staffetal2016MN8, NandezIvanova2016, Kuruwitaetal2016, IvanovaNandez2016}; for review on the CEE see, e.g., \citealt{Ivanovaetal2013}). Although in many cases the energy released by the in-spiraling binary system is more than what is required to unbind the CE (e.g., \citealt{DeMarcoetal2011, NordhausSpiegel2013}), the removal of the CE in numerical simulations encountered some problems  (e.g., \citealt{DeMarcoetal2011, Passyetal2011, Passyetal2012, RickerTaam2012, Ohlmannetal2016}).

Different extra energy sources have been suggested to facilitate the envelope removal, such as the recombination energy of the ejected envelope (e.g., \citealt{Nandezetal2015} for a recent paper).
\cite{Soker2004} suggests that in some cases, both for stellar and sub-stellar companions, no extra energy source is required, but rather the energy source might be the giant luminosity itself, namely, the nuclear burning in the giant core.  For that possibility to occur \cite{Soker2004} assumes that rapidly spinning envelopes have high mass-loss rates, e.g., by enhancing dust formation.

Relevant to this review is the suggestion that in many cases jets launched by the companion might  supply the extra energy required to expel the envelope (e.g., \citealt{Soker2014}). This holds for main sequence stars as well \citep{SchreierSoker2016}, as they can accrete mass at a high rate \citep{Shiberetal2016}. \cite{ArmitageLivio2000} and \cite{Chevalier2012} study CE ejection by jets launched from a NS companion, but they did not consider jets to be a general common envelope ejection process. In any case, a NS companion can launch very energetic jets when it enters the core of the giant star, and have an effect on the primary star that is similar in many respects to the explosion of single stars by jets. The JFM that operates during the spiraling-in of a NS in a giant envelope will leave a bare NS at the end \citep{Papishetal2015b}.
\cite{Papishetal2015b} further suggest that the jets launched by the NS as it enters the core of the giant star might be a site where  strong r-process nucleosynthesis, where elements with high atomic weight of $A > 130$ are formed, takes place.

\cite{Soker2004} considers white dwarfs companions that launch jets inside the envelope. White dwarfs that accrete mass at high rates sustain nuclear burning on their surface and expand (e.g. \citealt{Hachisuetal1999}). \cite{Soker2004} shows that under some conditions even in that case  jets will be launched by the inflated white dwarf envelope. The nuclear energy production on the surface of the white dwarf might boost the energy of the jets.

\cite{Soker2004} dismisses main sequence stellar companions for jet launching inside a giant envelope, because no Keplerian disk is expected to be formed around a main sequence star orbiting deep inside the envelope of a giant star. However, it seems now that even main sequence stars can launch jets inside the envelope.
Consider a main sequence star accreting mass inside the envelope of a giant star. In many cases the specific angular momentum of the accreted gas, $j_{\rm acc}$, is lower than the minimum value required to form a Keplerian accretion disk on the surface of the MS star, $j_{\rm Kep}$.
However, the ratio is still non-negligible with $j_{\rm acc} \approx 0.1j_{\rm Kep}- j_{\rm Kep}$ \citep{Soker2004}. In such a case an accretion belt is likely to be formed around the main sequence star, and jets might be launched by the accretion belt \citep{SchreierSoker2016}.
Without any cooling the high pressure that is built by the accreted gas might prevent the formation of an accretion disk or an accretion belt (e.g. \citealt{MacLeodRamirezRuiz2015}).
The removal of energy and high-entropy gas by the jets reduces the pressure near the accreting star, hence eases the accretion process and allows high accretion rates \citep{Shiberetal2016, Staffetal2016MN}, and the formation of an accretion disk or belt.
{{{{ \cite{RickerTaam2012}, for example, find the mass accretion rate onto the secondary star in their simulation of the CEE to be much below the one given by the Bondi-Hoyle-Lyttleton (BHL) prescription. However, they do not include jets that can remove angular momentum and high-entropy gas from the vicinity of the secondary star. It is expected that when jets are launched and remove angular momentum and high-entropy gas, }}}} the accretion rate will be closer to the Bondi-Hoyle-Lyttleton prescription, hence much higher than that obtained in the CEE numerical simulation of \cite{RickerTaam2012}.
{{{{ Still, some assumptions of the BHL flow do not apply in the CEE \citep{RickerTaam2012}, and the accretion rate will be somewhat lower than that given by the BHL rate, even when jets are launched. }}}}
The mass accreting star can then launch $10-40\%$ of the accreted mass into the jets, as it is found for jets from YSOs (e.g., \citealt{Pudritzetal2012, Federrathetal2014}).

As with other cases of the JFM, the importance of jets stem from the deep potential well of the accreting object relative to that of the ambient gas. The energy that is carried by the jets can be compared to the energy released by the spiraling-in process as given by the $\alpha_{\rm CE}$-prescription (see also \citealt{Soker2014}).
The energy released by an accreting MS star is
\begin{equation}
E_{\rm jets} \simeq \frac {G M_2 M_{\rm acc}}{2R_2} ,
  \label{eq:jete1}
\end{equation}
where $M_2$ is the mass of the secondary main sequence star, $R_2$ is its radius, and $M_{\rm acc}$ is the mass it accretes. The energy released by the in-spiraling main sequence star that is channelled to envelope removal is
\begin{equation}
E_{\alpha_{\rm CE}} = \frac {G M_{\rm core}{M_2}}{2a_{\rm final}}
\alpha_{\rm CE},
  \label{eq:CEalpha1}
\end{equation}
where $a_{\rm final}$ is the final orbital separation of the core, of mass $M_{\rm core}$, and the companion. For the accretion process onto the main sequence star to release energy about equals to that released in the $\alpha_{\rm CE}$-prescription, the accreted mass onto the secondary star must be
\begin{equation}
M_{\rm acc-\alpha} \simeq 0.06 \left( \frac{M_{\rm core}}{0.6
M_\sun} \right) \left( \frac{R_2}{1 R_\sun} \right) \left(
\frac{a_{\rm final}}{5 R_\sun} \right)^{-1} \left(
\frac{\alpha_{\rm CE}}{0.5} \right) M_\sun.
  \label{eq:accm1}
\end{equation}

Namely, jets launched by a secondary star that accretes $\approx 10 \%$ of its mass can play a significant, and even the major, role in removing the giant envelope. Moreover, the jets can ejects large parts of the envelope at very high velocities, i.e., much above the escape speed from AGB stars \citep{Soker2015a}

The JFM operates as follows. If jets are indeed launched, they inflate bubbles and deposit energy to the common envelope. The deposited energy removes some mass from the envelope and causes the envelope to expand. Both processes reduce the average mass density in the vicinity of the companion, hence reducing the accretion rate. This cycle operates in a negative feedback.

There might be also a positive effect. As the jets remove mass mainly from the polar directions, most of the mass accreted by the companion comes now from near the equatorial plane. This mass has a higher specific angular momentum than the mass that potentially could have been accreted from the polar directions. The formation of an accretion disk becomes more likely. However, it is possible that the accretion belt itself can efficiently launch jets, such that the formation of an accretion disk will not change much.  The requirement that accretion belts around main sequence stars launch jets is similar to the requirement from accretion belts around NS in the jittering-jets model for the explosion of massive stars (section \ref{subsec:SNe:belt}).

The spiraling-in process leads to a flatten envelope outside the orbit of the companion.
As discussed in section \ref{sec:PNe}, jets can further compress gas toward the equatorial plane.
A circumbinary disk is formed around the core-companion system.
It is possible that in some cases at late times the mass inner to the orbit of the core and the companion is low, but mass in the circumbinary disk is large.
\cite{Kashisoker2011MN} suggest that the interaction of the circumbinary disc with the binary system might reduce the orbital separation much more than expected during the preceding dynamical phase, when part of the envelope was ejected and the circumbinary disk was formed.
This situation resembles that of planet-migration in protoplanetary disks. Extra energy can be carried out by the efficient convection in the envelope, and then radiated away \citep{Soker2014}.

A very complicated structure of jets and outflows might arise when the giant star engulfs a binary system instead of a single companion. In that case each of the two stars in the engulfed tight binary system might launch jets, either simultaneously, or each one on its turn \citep{SabachSoker2015, Soker2016a}.
Furthermore, the two stars of the tight binary system might merge inside the giant envelope. An accretion disk might form around the surviving remnant, and launch jets. Such energetic jets might remove a large fraction of the envelope in a highly-asymmetrical outflow.
This process is expected to form PNe with `messy' morphologies, namely, lack any symmetry \citep{Soker2016a}.
In another outcome of the triple-stellar CEE the tight binary system breaks-up inside the envelope \citep{SabachSoker2015}. Jets can be launched by the stars in this type of interaction as well.

In Table 3 I list the possible stages of the CEE relevant to the topic of this review, and the possible roles jets might play.
\begin{table}[!htb]
\label{Tab:TableCE}
\begin{tabular}{|l|l|l|l|l|}
\hline
\small {Secondary star}&{Source of}    &{Role of Jets}&{Observational}& {Comments} \\
\small {and its orbit }&{accreted mass}&              &{signatures}   &          \\
\hline
\small (a) Outside envelope;   & RLOF         & Shaping the slow  & Bipolar symbiotic   & No JFM  \\
\small mainly in           &              & giant wind        & nebulae; Bipolar PNe   &  \\
\small synchronization    &              &                    & with a narrow waist    &  \\
\hline
\small (b) Spiraling-in    & BHL          & Removing          &           & Possible jets  \\      \small in outer CE         & accretion    & part of the       &           & from accretion  \\
\small                     &              & envelope          &           & belts; JFM    \\
\hline
\small (c) Inner CE:       & Circumbinary & Removing and       & Elliptical& JFM  \\
\small migrates-in due to  & thick disk   & accelerating       & structure&     \\
\small circumbinary disk   & (flatten CE) & the CE             & in the PN&   \\
\hline
\small (d) Post CE:       &Circumbinary &Shaping the nebula;&Elliptical PNe+ansae;&Jets might be\\
\small Residual migration &thick disk or&Forming hot        &Mildly bipolar PNe.&lunched also\\
\small                    &fall-back gas&bubbles in         & Diffuse X-ray in& from the core \\
\small                    &             &elliptical PNe     & elliptical PNe   &           \\
\hline
\small (e) Merger (during &Destroyed   & Shaping the nebula;& Elliptical PNe+ansae;& Jets might\\
\small or after CE ejection).&secondary or& Forming hot        & Mildly bipolar PNe.& be highly \\
\small Core launches jets &fall-back gas& bubbles in         & Diffuse X-ray in & collimated \\
\small                     &              &  elliptical PNe    & elliptical PNe       &    \\
\hline
\small (f) Grazing the    & RLOF +       & Removing entire & Bipolar nebula; & Very efficient\\
\small envelope: GEE       & BHL       & envelope        & possibly large  & JFM \\
\small                     &       & outside orbit   & final orbit     &  \\
\hline
\end{tabular}
 \footnotesize
\newline
\caption{
Possible stages of the CEE of a late AGB primary star and a main sequence secondary star (companion). Many items are relevant to other types of binary systems as well.
This is an updated version of a table presented by \cite{Soker2014}.  `Ansae' stand for two opposite small bullets, one at each side of an elliptical PN, that generally move faster than the rest of the nebula.
BHL: Bondi-Hoyle-Lyttleton type accretion.   RLOF: Roche lobe overflow accretion.
}
\end{table}

\textbf{$\bigstar$ Summary of section \ref{sec:CEE}.} I summarize this section by emphasizing the following processes, some of which are still speculative and require a great deal of further work.
\newline
(1) Main sequence stars, white dwarfs, and  NS companions spiraling-in inside the envelope of a giant star might accreted mass at a high rate and launch jets that facilitate envelope removal.
\newline
(2)
In the case of a main sequence companion in many cases the accretion flow might be sub-Keplerian, and an accretion belt is formed rather than an accretion disk (at least initially). This further raises the possibility that accretion belts can efficiently launch jets, maybe even more efficiently than accretion disks do. The launching of jets by an accretion belt must be studied in greater details.
\newline
(3) The spinning-up of the envelope and the inflation of bubbles by jet can lead to the formation of a relatively massive circumbinary thick disk (flatten envelope). The circumbinary disk can lead to migration, i.e., the shrinkage of the orbital separation. The released gravitational energy is channelled both to further mass loss, but also in large part to radiation.
\newline
(4) In cases where envelope removal is not efficient, and in particular when migration takes place, the companion and the core might merge.

\section{Grazing Envelope Evolution (GEE)}
\label{sec:GEE}

The GEE was proposed very recently \citep{Soker2015a} and it is still in a speculative phase of preliminary development.
If jets are launched by the companion and become efficient in removing the envelope when the companion just enters the giant envelope, such that the jets and the bubbles they inflate remove the entire envelope outside the orbit of the companion, the system does not enter a CEE. Instead, the system might experience a GEE (\citealt{SabachSoker2015, Soker2015a, Soker2016a, Soker2016b}). The GEE can be viewed as a continues process of `just-entering' the CEE.

The basic ingredients and properties of the GEE are proposed to be as follows.
\newline
(1) Jets launched by the secondary star remove the entire envelope outside the orbit of the secondary star.
  \newline
(2) As the secondary touches the envelope, the mass transfer process is a hybrid of the Roche lobe overflow (RLOF) accretion and the Bondi-Hoyle-Lyttleton (BHL) type accretion.
  \newline
(2) The secondary star cannot accrete more than the mass in its vicinity, hence the jets limit the mass available for accretion. Namely, the operation of the jets is regulated by a negative feedback mechanism, the JFM. Unlike in the CEE, the GEE requires the operation of a JFM.
  \newline
(3) To have sufficiently strong jets, the companion must accrete at a high rate. Recent results support the notion that main sequence stars can accrete mass at a high rate through an accretion disk or an accretion belt \citep{Shiberetal2016, Staffetal2016MN}.
The 3-dimensional hydrodynamical simulations conducted by \cite{Staffetal2016MN} are very constructive in demonstrating that a main sequence companion can indeed accrete mass at a high rate, $\approx 0.01 M_\odot \yr^{-1}$, and might form a thick accretion disk when orbiting close to and outside a giant envelope. This is exactly the accretion flow that is required by the JFM that gives rise to the GEE.
  \newline
(4) The spiraling-in process is mainly caused by the interaction of the companion with the envelope. As it touches the envelope, the interaction is complicated to calculate. It is a hybrid of a tidal interaction and a gravitational-drag interaction. Accretion further acts to reduce the orbital separation, while the mass ejected by the jets acts to increase the orbital separation.
Mass that might be lost through the second Lagrangian point (beyond the secondary star) can carry additional angular momentum and envelope mass.
  \newline
(5) In cases where the secondary star is relatively light, the tidal forces will cause it to spiral-in. In those cases the GEE can serve as an alternative to the CEE in shrinking the orbit.
  \newline
(6) In some other cases, when the secondary is not too light and the jets are effective in removing the envelope, the orbital separation stays large, and can even increase.
\newline
(7) The interaction of the jets with gas that was ejected in previous orbits, in the stellar vicinity, channelled kinetic energy to radiation. This might lead to an intermediate luminosity optical transient (ILOT; this is discussed in section \ref{sec:ILOTs}).

There are post-AGB binary systems with a relatively large orbital separation and with a nebula around them that have been shaped by jets. The possibility that these systems have evolved through the GEE should be examined in future studies \citep{Soker2016b}.
One example is the post-AGB \astrobj{Red Rectangle} bipolar nebula (e.g., \citealt{VanWinckel2014} for its properties). As a side comment, and in relation to the usage of wide jets in cooling flows and other systems, I note that the jets launched by the main sequence companion to the post-AGB star in the Red Rectangle central binary system have a wide opening angle of $\alpha_j \simeq 35^\circ$ \citep{Thomasetal2013}.
Other examples for post-AGB binary systems that potentially experienced GEE might be \astrobj{BD$+46^\circ442$} that has an orbital period of $140.77~$days and was found to launch jets \citep{Gorlovaetal2012}, and \astrobj{IRAS~19135+3937} with an orbital period of 127 days and a collimated outflow blown by its companion  \citep{Gorlovaetal2015}.

\textbf{$\bigstar$  Summary of section \ref{sec:GEE}.} This short section can be summarized with one significant point. If the JFM does indeed operate in some cases of CEE, a very efficient JFM might prevent the CEE all together. Instead, a GEE takes place. In cases of a main sequence secondary star, it is required that main sequence stars could be able to accrete mass at a high rate and launch jets.

\section{Intermediate-luminosity optical transients (ILOTs)}
\label{sec:ILOTs}

The suggestion that in some ILOTs the JFM operates to some degree is very new \citep{KashiSoker2016}, and like the GEE it is still in a speculative and preliminary stage of development. Unlike the GEE where the JFM is required, the JFM is not necessary for the operation of jets in ILOTs; it was basically discussed only in the above paper.

Not only that there is no consensus yet on the exact powering mechanism of ILOTs, there is not even a consensus on the terminology.
According to the terminology suggested by \cite{KashiSoker2016}, the group of ILOTs is a heterogeneous group that includes the following subgroups.
\newline
\textbf{ILRT: Intermediate-Luminous Red Transients}. These are ILOTs in evolved stars, such as AGB stars. It might be that a main sequence companion accretes mass and power the ILOT.
\newline
\textbf{LBV giant eruptions and SN Impostors}. Giant eruptions of
Luminous Blue Variables. Examples include the 1837--1856 Great Eruption f
$\eta$ Carina and the pre-explosion eruptions of \astrobj{SN~2009ip}.
ILRTs are the low mass relatives of LBV giant eruptions.
\newline
\textbf{LRN or RT: Luminous Red Novae or Red Transients or
Merger-bursts.} These are powered by a full merger of two stars.
Merger events of stars with sub-stellar objects are also included.

Beside the possible operation of the JFM, some ILOTs are related to some systems that were discussed in previous sections, the CEE, the GEE, and PNe.
\cite{RetterMarom2003} mention the connection between the ILOT \astrobj{V838~Mon} and a CEE.
They suggest that the ILOT \astrobj{V838~Mon} was powered when a star swallowed three planets, one after the other \citep{Retteretal2006}.
The attribution of ILOTs to the CEE (or the onset of the CEE) but with a stellar companion followed with the ILOTs \astrobj{OGLE-2002-BLG-360} \citep{Tylendaetal2013}, \astrobj{V1309~Sco} \citep{Tylendaetal2011, Ivanovaetal2013a, Nandezetal2014, Kaminskietal2015}, and \astrobj{M31LRN~2015} \citep{MacLeodetal2016}. \cite{Pejchaetal2016a} and \cite{Pejchaetal2016b} study mass loss through the second Lagrangian point $L_2$ or during a CEE, and found that about $5 \%$ of the kinetic energy is thermalized in the collision of equatorially ejected gas. None of the above papers attribute any role to jets that might be launched in the different processes.

\cite{Soker2016b} raises the possibility that a large fraction of the kinetic energy carried by the jets during the GEE might be channelled to radiation, hence leading to an ILOT event.

\cite{SokerKashi2012} propose that the lobes of some (but not all) bipolar PNe and pre-PNe were formed in an ILOT event, or a series of several close ILOT events. They further suggest that in a several months long outburst a main-sequence companion accretes mass from an AGB star and launches jets, that both shape the bipolar PN and powered an ILOT event. The ILOT scenario was mentioned in relation to more PNe (\citealt{FrewParker2012, BoumisMeaburn2013, Clyneetal2014, Frewetal2016}).
\cite{Corradietal2014} raise the possibility that the YSO bipolar nebula \astrobj{Ou4} was formed in an ILOT event, as \cite{BoumisMeaburn2013} suggest for the PN \astrobj{KjPn~8} that has a similar morphology.

\cite{KashiSoker2016} adopt the High-Accretion-Powered ILOT (HAPI) model, according to which many (most) ILOTs are powered by a high-accretion rate event on to a main sequence, or slightly involved off the main sequence, star in a binary system \citep{KashiSoker2010arx}. The main sequence star launches jets that collide with the ambient gas, a process through which some kinetic energy is transferred to radiation. The reservoir of the accreted mass might be a giant star, such as an AGB star, an LBV star, like in the binary model of the great eruption of Eta Carinae \citep{Kashisoker2010Car}, or a destructed companion such as in the merger model for the ILOT \astrobj{V838~Mon} \citep{SokerTylenda2003}.

\cite{KashiSoker2016} discuss the possibility that the two opposite jets might expel a non-negligible amount of gas from this reservoir. By that the jets reduce and/or shorten, or even shut off, the mass accretion process. Namely, a JFM might take place in ILOTs.
\cite{KashiSoker2016} discuss three types of JFM situations, differ mainly in the duration of the accretion process, and in the mass reservoir.
\newline
(1) \emph{An ILOT event at a periastron passage.} A main sequence companion orbits a giant star, and the accretion phase lasts for a fraction of the orbit near periastron.
If jets are launched, they might reduce the accretion rate, but they cannot destroy the entire mass reservoir. If the companion survives the periastron passage the outburst might repeat itself. The best example is the Great Eruption of $\eta$ Carina, where at least two outbursts separated by about $5.2$ years have been identified \citep{Damineli1996, SmithFrew2011}.
\newline
(2) \emph{A fall back gas.} In this more speculative case, the JFM takes place at late times after the outburst was triggered. The accreted gas comes from mass that was ejected in a violent event, like a merger event. Part of the ejected gas might fall back to the remnant. The fall back gas might form an accretion disk around the merger remnant, that in turn might launch jets. These jets might expel some gas from the reservoir, in a JFM process.
\newline
(3) \emph{ILOT in a GEE}. This was discussed in section \ref{sec:GEE}.

Last word about accretion. Three-dimensional hydrodynamical simulations show that if the secondary star in the binary system Eta Carinae accretes mass at each periastron passage, the accretion is of dense clumps  \citep{Akashietal2013, Kashi2016}. The dense clumps can penetrate through the secondary wind. This might be the case in many ILOTs. Hence, ILOTs are another type of objects where accretion onto the compact object that launch jets is via dense clumps, rather than of a smooth medium. The other objects are cooling flows and young galaxies. It should be examined whether in GEE and CEE accretion of dense clumps also takes place.

\textbf{$\bigstar$ Summary of section \ref{sec:ILOTs}.} I summarize this section by stating again that the possibility that a JFM takes place in some ILOTs is speculative.  More than that. Because the typical ratio of gravitational potential on the mass accreting body to that in the reservoir $\Phi_a/\Phi_{\rm res} \la 10$ in merger events is relatively small, the JFM in ILOTs that result from a merger process is not very efficient.
Nonetheless, as ILOTs might occur along side the CEE, the GEE, and the formation process of some bipolar PNe, it is important to further study this speculative suggestion.

\section{Young Stellar Objects (YSOs)}
\label{sec:YSOs}

Feedback in star forming clouds is reviewed by, e.g., \cite{Vazquez2011}, \cite{Franketal2014},   \cite{Krumholzetal2014}, and \cite{Dale2015}. The feedback during star cluster formation is likely to explain the inefficient process of star formation (e.g. \citealt{Federrath2016}), it might explain why stars disperse from their birth sites within a times scale of $\ll 1 \Gyr$,
and it plays a role in determining the initial stellar mass function \citep{Krumholzetal2014}.

Despite the spectacular jets that are observed expanding from many young stellar objects (YSOs; e.g., \citealt{ReipurthBally2001}) that in some cases reveal an hourglass morphology (e.g., \citealt{Stephensetal2013}) and lobes (e.g., \citealt{Purseretal2016}), the JFM in YSOs is not as efficient as in some other astrophysical systems. The reason is that in most cases jets from young YSOs do not obey one or both of the conditions for the efficient operation of the JFM. In many cases jets from YSOs are narrow and well collimated, expanding to very large distances and form Herbi-Haro objects (e.g. \citealt{ReipurthBally2001}). These jets do not influence much the formation of their parent star.
As well, the shocks that are formed when jets from young stellar objects interact with the cloud are highly radiative. This implies that most of the kinetic energy of the jets is lost to radiation, and to a good approximation the jets may be considered as sources of momentum only, i.e., are of the momentum-conserving case \citep{Krumholzetal2014}.
When hot massive stars are presence, ionizing radiation, radiation pressure, and stellar winds play a larger role than jets launched by YSOs in the feedback mechanism in star forming clouds (e.g., \citealt{Smithetal2010, Cunninghametal2011, Krumholzetal2014}). When only low and medium mass stars are formed, jets have more pronounced role.

Even in the momentum-conserving case the JFM might be significant to the cloud evolution, in particular if jets from several stars act together \citep{Petersetal2014}.
The jets, even from low mass stars, have enough momentum to maintain a supersonic turbulence in
clumps from which they were formed {{{{ (e.g., \cite{Carrolletal2010}), }}}} and even eject some of the mass in the clump, hence negatively influencing further star formation (e.g., \citealt{Nakamura2016}).

Another process by which jets play a role is by opening low-optical depth regions along the polar direction. \cite{Cunninghametal2011} find that the radiation from the young star can escape more freely along the polar cavities that are opened by the jets, hence reducing the radiation pressure in the equatorial plane.
Jets can remove some mass from the cloud in which the stars are forming, and can increase somewhat the turbulence in the cloud (e.g., \citealt{Franketal2014} and references therein).
As with the process of bubble-inflation, a relative transverse motion between the jets and the ambient gas increases the efficiency of momentum transfer from the jets to the ambient gas \citep{Franketal2014}.

\cite{Federrathetal2014} simulated the feedback from jets and outflows in star cluster formation. Their main results are that jets eject about one-fourth of their parent molecular clump to a distance of $\approx 1 \pc$, and the jets reduce the star formation rate by about a factor of two.

There are some basic differences between the JFM operating in YSOs and the other types of objects discussed here. (1) In most cases the jets of one star influence a cloud (or a clump) that feeds many stars. I term this JFM a \emph{collective JFM}. (2) In most cases the jets lose most of their kinetic energy to radiation after they experience a shock wave. They transfer momentum to their ambient gas, e.g., momentum-conserving case. (3) In most cases jets are not the dominant feedback process. There are ionizing radiation and stellar winds that are powered by nuclear burning rather than accretion.

There are also some common properties. Some YSOs show slow-massive-wide (SMW) bipolar outflows, e.g., the molecular outflow reported by \cite{Higuchietal2015}. Such an outflow might be similar in some respects to the SMW outflows reported from AGB (section \ref{sec:CFs}).
The collective JFM can lead to results similar to those from a single accreting objects in other types of systems. \cite{Liuetal2016} discuss a very slow and extremely wide outflow formed by jets in a momentum feedback process in a clump of a mass of $\approx 210 M_\odot$, that might be forming several protostars. The formation of a slow wide outflow might be similar in some respect to the slow wide outflows found in some AGN (e.g., \citealt{Aravetal2013}).

So why is the JFM works for YSOs despite the fast radiative cooling, such that the jets deposit their momentum rather than initial kinetic energy? One can see this from the ratio   $\Phi_a/\Phi_{\rm res} \approx 10^5$ as given in Table 1. It is much higher than in other stellar objects. It has a value similar to that in galaxy formation and cooling flows. However, in the later two the mass of the SMBH is $\la 10^{-3}$ that of the ambient gas. The combined mass of the stars is  $\approx 0.01$ times the mass of the clump from which they have been formed. A large fraction of the stellar mass is accreted through an accretion disk that launch jets.
Due to the large potential ratio and the not-too-small stellar to clump mass ratio, the momentum of the outflows from YSOs can significantly influence their parent clump.

At very early phases of star formation the accretion rate might be very high, and the gas optically thick. This implies that radiation cooling might take longer, and bubbles can be formed close to the star. In that case the feedback might work in the energy-conserving case and be more effective.

\textbf{$\bigstar$ Summary of section \ref{sec:YSOs}.} Although winds, radiation, and SN explosions of massive stars deposit more energy to the parent cloud than jets do, the jets nonetheless still play a role. Jets from a star might influence to some extend the immediate vicinity of the star by ejecting some mass and compressing gas to the equatorial plane (if bubbles are inflated).
They are more likely to influence the clump (cloud) at larger distances, in particular if jets from several stars act together in a collective JFM. This process reduces star formation rate. In young galaxies the stellar JFM that operates in the cloud might operate while an AGN-driven JFM operates in the entire galaxy. Namely, we have a JFM within a JFM.

\section{Summary of open questions}
\label{sec:summary}

The structure of this review is such that a short summary is given at the end of each section (or at the end of each subsection in section \ref{sec:CFs}), and the properties of the different groups are summarized in Table \ref{table:compare}. I therefore will not summarize the main points in this section. I rather list some open questions, and my view on the most likely answers to the questions.

\emph{(1) Does a JFM mechanism operate in CCSNe, CEE, GEE, and ILOTs? } In these systems the suggestion that a JFM takes place is speculative and controversial. In particular in CCSNe. I hope the comparison of the eight different types of objects studied here, where in some the JFM is known to take place, will encourage further studies of the possibility that a JFM does take place in CCSNe, in the CEE, in the GEE, and in ILOTs.
My view is that a JFM does operate in many CCSNe, in some cases of CEE, in all cases of GEE (that are made possible by jets), and in some ILOTs. I argue that jets are much more common in astronomical objects than what is inferred from objects where they are directly observed.

\emph{(2) What is the heating mechanism in cooling flows? } As cooling flows are the only objects  where the full JFM cycle can be observed, it is important to better understand its operation.
One open question is how energy is transferred from the jets to the ambient gas. Several processes have been suggested. My view is that mixing of the hot gas in jet-inflated bubbles with the ambient gas heats the ambient gas.

\emph{(3) To what extended the AGN-JFM play a role in determining correlations of properties in galaxies?}  There is a question whether the AGN-JFM determines the correlations between the mass of the SMBH and some properties of the stellar content of elliptical galaxies, or bulges of spiral galaxies. If it does, then the question is what is the exact AGN-JFM mechanism. For example, does it operate in a momentum conserving or an energy-conserving interaction. My view is that an energy-conserving AGN-JFM is behind these correlations.

\emph{(4) Can accretion belts launch jets?} In some cases the accreted gas has a specific angular momentum that is below the value that is required to form a Keplerian accretion disk, but it is still large enough to form an accretion belt, i.e., $j_{\rm acc} \approx 0.1j_{\rm Kep}- j_{\rm Kep}$. Can such accretion belts launch energetic jets? My view is that they can. But this question really deserves much deeper studies.

\emph{(5) What is the outcome of the failure of a JFM?} In Table \ref{table:compare} I list the outcome when the JFM fizzles or fails. According to the perspective of this review, some very interesting and energetic events owe their existence to the failure of the JFM, including stellar black holes, gamma ray bursts, and type Ia supernovae. But there is much more to explore about the fizzle of the JFM.

{{{{ I thank the referee Eric Blackman for many helpful comments. }}}}
This research was supported by the E. and J. Bishop Research Fund at the Technion.

\end{document}